\documentclass[floats,floatfix,showpacs,amssymb,prd,twocolumn,superscriptaddress,nofootinbib,nolongbibliography,reprint]{revtex4-2}

\usepackage{amssymb,amsmath,verbatim,mathtools,needspace,enumitem,etoolbox,graphicx,physics,microtype,afterpage,bigints,gensymb,tabularx,xspace}

\usepackage[dvipsnames, usenames]{xcolor}
\definecolor{linkcolor}{rgb}{0.0,0.3,0.5}
\definecolor{dodgerblue}{HTML}{1E90FF}
\usepackage[unicode, colorlinks=true, linkcolor=linkcolor, citecolor=linkcolor, filecolor=linkcolor,urlcolor=linkcolor, pdfusetitle]{hyperref}
\usepackage[all]{hypcap}
\usepackage[T1]{fontenc}
\usepackage[utf8]{inputenc}
\usepackage{orcidlink}

\renewcommand{\vec}[1]{\mathbf{#1}}

\makeatletter
\newcommand*{\balancecolsandclearpage}{\close@column@grid \cleardoublepage \twocolumngrid}
\makeatother

\newcommand{\milan}{\affiliation{Dipartimento di Fisica ``G. Occhialini'', Universit\'a degli Studi di Milano-Bicocca, Piazza della Scienza 3, 20126 Milano, Italy}}
\newcommand{\infn}{\affiliation{INFN, Sezione di Milano-Bicocca, Piazza della Scienza 3, 20126 Milano, Italy}}

\usepackage[normalem]{ulem}

\begin{document}

\title{Not all spins of black-hole binaries formed in clusters are isotropic \\ and not all spins of black-hole binaries formed in isolation are aligned}

\author{Sofia Dossena$\,$\orcidlink{0009-0008-0928-1116}}
\email{s.dossena8@campus.unimib.it}
\milan

\author{Davide Gerosa$\,$\orcidlink{0000-0002-0933-3579}}
\milan \infn

\author{Tristan Bruel$\,$\orcidlink{0000-0002-1789-7876}}
\milan \infn


\date{\today}

\begin{abstract}

Spin directions are promising observables for distinguishing between the formation channels of merging stellar-mass binary black holes with gravitational-wave observations. In this work, we challenge the standard expectation that binaries formed via dynamical interactions in stellar clusters have isotropically distributed spins and that binaries formed in isolation have spins aligned with the orbital angular momentum. For dynamically formed binaries, we account for observational findings suggesting that the spins of stars in clusters may exhibit a preferred degree of alignment and construct a simple geometrical description of the resulting black-hole spin directions. This results in an analytical joint distribution for the polar angles of first-generation black-hole binaries, whose key feature is a correlation between the two spin-orbit tilts. We show that, under reasonable assumptions and independently of the initial spin configuration, spins are fully randomized after a single black-hole merger, implying that binaries formed through hierarchical mergers have isotropically distributed spin orientations. For isolated binaries, we introduce a simple model for spin evolution as a function of the natal kick imparted during the first supernova explosion, while accounting for tidal interactions and mass-ratio reversal. The resulting distribution is also fully analytic. We investigate the impact of relativistic spin precession, as well as that of the finite escape speed of clusters, on our geometrically derived distributions. Our results provide a simple analytical framework for modeling the spin directions of merging black holes in gravitational-wave population fits.

\end{abstract}

\maketitle

\section{Introduction}
Despite the growing number of gravitational-wave (GW) detections~\cite{2026arXiv260527226T}, understanding the astrophysical origin of merging stellar-mass binary black holes (BHs)—the so-called formation-channel problem—remains an open issue. Astrophysical scenarios proposed to explain the observed demographics of GW events can be broadly divided into two main classes: dynamical formation in dense stellar environments, such as star clusters, and isolated binary stellar evolution in the galactic fields~\cite{2022PhR...955....1M, 2021hgwa.bookE..16M}.

In the former scenario, mergers are driven by gravitational multi-body interactions. A distinctive outcome of the dynamical channel is the possibility of hierarchical mergers, which occur when the remnant of a previous BH merger is retained within the cluster and subsequently undergoes one or more additional mergers~\cite{2021NatAs...5..749G}.

In the latter scenario, GW sources originate from pairs of massive stars that are born together and evolve in isolation. Stellar evolutionary processes such as mass-transfer phases are typically invoked to decrease the orbital separation till the final GW-driven regime. %

Possible observables for distinguishing between the two formation channels include the BH masses. While newer datasets are revealing increasingly fine structure in the merging BH mass spectrum~\cite{2026arXiv260527226T}, their astrophysical interpretation is somewhat challenging because the mass distributions predicted by the dynamical and isolated channels might overlap substantially (e.g.~\cite{2019ApJ...886...25B,2021ApJ...910..152Z}). %
BH spin magnitudes are also not particularly effective at distinguishing between formation channels because they are largely determined by the physics of stellar collapse~\cite{2011ApJ...730...70O,2019ApJ...881L...1F}, which is common to both channels (hierachical mergers, which only occur in dynamical settings, are a notable exception). Orbital eccentricity is essentially exclusive to the dynamical formation channels, but confident detections are still elusive.

In this context, spin orientations have been recognized as one of the most discriminating observables since the early days of observational GW astronomy~\cite{2016ApJ...832L...2R,2017MNRAS.471.2801S,2017CQGra..34cLT01V,2017PhRvD..96b3012T,2018PhRvD..98h4036G,2023ApJ...946...50B}. The key idea is that, in the isolated channel, the stellar binary progenitors form from the same gaseous environment, inheriting a common angular momentum and therefore producing BHs whose spins are expected to be largely aligned with the binary’s orbital angular momentum. %
In contrast, binaries assembled dynamically undergo multiple gravitational encounters that effectively erase memory of their formation conditions. 
Overall, the broad prediction is that the spins of merging BHs formed in isolation should be  aligned with the binary’s orbital angular momentum, whereas the spins of BHs formed dynamically should be isotropically distributed.

While this represents the zeroth-order picture, several phenomena can modify spin orientations in both channels. In the field scenario, supernova (SN) kicks~\cite{2000ApJ...541..319K}, tidal interactions~\cite{2020A&A...635A..97B}, and mass transfer~\cite{2021PhRvD.103f3007S,2026arXiv260811311M} during stellar evolution, as well as relativistic spin couplings during the early inspiral~\cite{2013PhRvD..87j4028G}, can either introduce or suppress spin-orbit misalignments (see e.g. Refs.~\cite{2018PhRvD..98h4036G,2021PhRvD.103f3032S,2024arXiv241203461B} for end-to-end predictions). In the dynamical case, on the other hand, there is observational evidence from asteroseismology that the spin directions of stars in clusters exhibit some degree of mutual alignment~\cite{2017NatAs...1E..64C} (but see e.g.~\cite{2023ApJ...944...39H} for a contrasted view on this result), and whether any memory of this correlation is retained in merging BHs has not been thoroughly explored.
 
Building on these expectations, we present an analytical model for the spin directions of merging binary BHs in both the field and cluster scenarios. Our goal is to describe deviations from full alignment and full isotropy, respectively, in terms of astrophysical parameters. We then test its applicability by incorporating the effects of post-Newtonian (PN) evolution between BH binary formation and merger, as well as the finite escape velocity of clusters. 
The result is a prescriptive model designed for use in GW population fits.

\section{Not all spins of BH binaries formed in clusters are isotropic} \label{clustersec}

In this section, we present our model for the spin orientations of BH binaries formed in clusters, relaxing the common assumption of perfect isotropy. Some mathematical derivations are postponed to Appendices~\ref{label:jointcluster} and~\ref{tossapp}.

\subsection{Geometric description}
\label{sec:geometry}

\begin{figure*}
    \centering
    \includegraphics[width=0.8\textwidth]{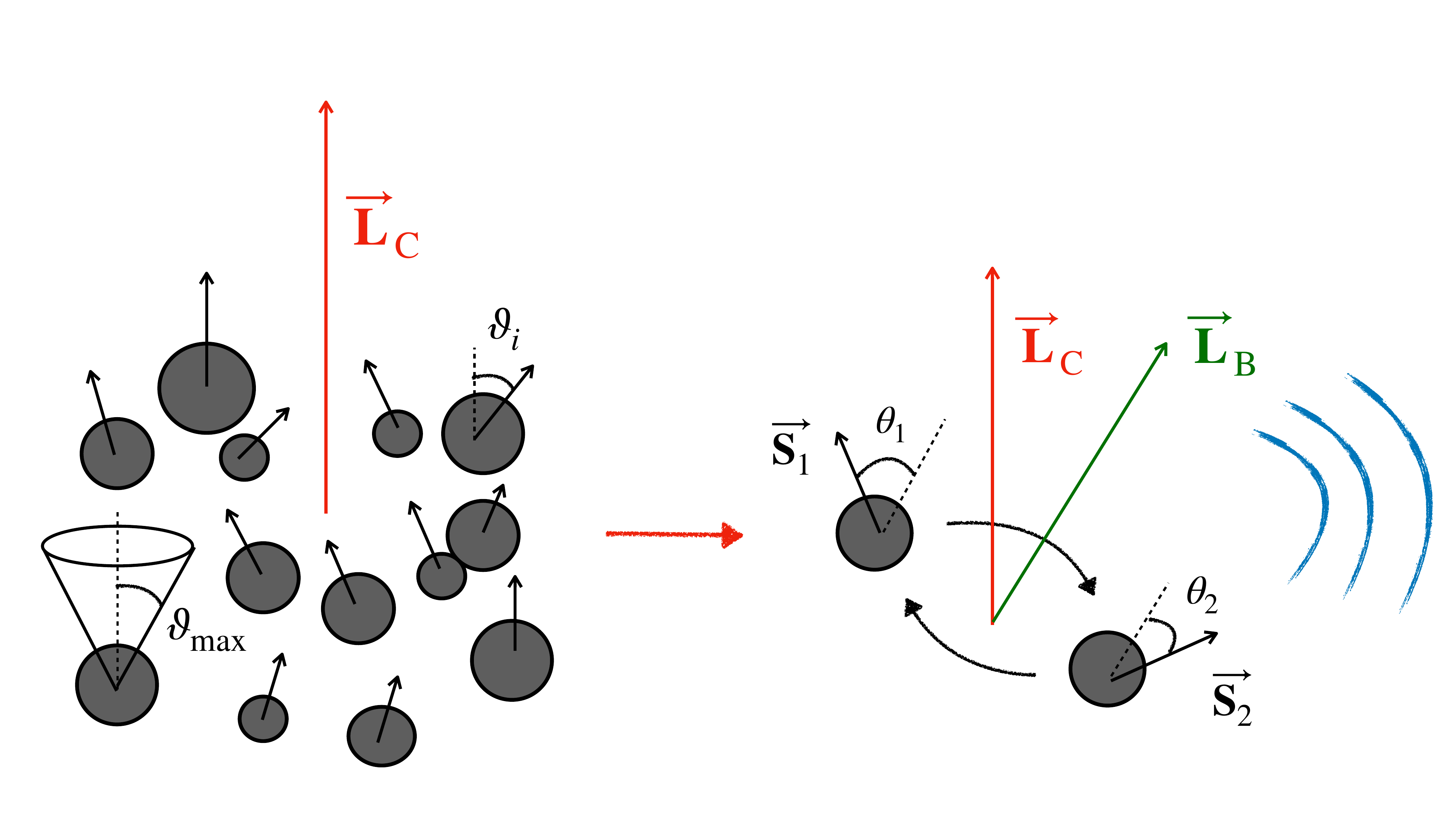}
    \caption{%
    Schematic representation of our model for BH binary spin directions in stellar clusters. The cluster angular momentum is denoted by $\vec{L}_{\rm C}$, and the BH spin directions are assigned by drawing the polar angle of the $i$-th BH from $\cos\vartheta_i\sim \mathcal{U}[\cos\vartheta_{\mathrm{max}},1]$ (left). The binary orbital angular momentum $\vec{L}_{\rm B}$ is randomly oriented as a result of dynamical interactions. The spin tilt angles $\theta_1$ and $\theta_2$ are then defined as the angles between $\vec{L}_{\rm B}$ and the spin directions $\vec{S}_1$ and $\vec{S}_2$, respectively (right).}
    \label{fig:fig1}
\end{figure*}

Figure \ref{fig:fig1} shows a schematic representation of the cluster geometry and spin directions at BH binary formation.   
Let us denote the angular momentum of the cluster by $\vec{L}_{\rm C}$; this sets an overall preferential direction. Deviations from isotropy %
are controlled by a global parameter $\vartheta_{\text{max}}$.

We specify each BH spin direction relative to $\vec{L}_{\rm C}$ using a polar angle $\vartheta$ and an azimuthal angle $\varphi$, drawn from $\cos\vartheta\sim \mathcal{U}[\cos\vartheta_{\mathrm{max}},1]$ and $\varphi\sim \mathcal{U}[0,2\pi]$, respectively.
Models with $\vartheta_{\text{max}}\sim 0$ correspond to the case in which the spins of BH binaries ``remember'' their correlated stellar formation in the same gaseous environment~\cite{2017NatAs...1E..64C}. The limiting model with $\vartheta_{\text{max}}= \pi$ corresponds to full isotropy, as commonly assumed in GW astronomy.

BH binaries in clusters are formed and hardened by multiple few-body interactions. We assume these cause the binary's orbital angular momentum $\vec{L}_{\rm{B}}$ to be randomly oriented (see e.g. Ref.~\cite{2026arXiv260621691M} for detailed models).  
The relative orientations of the two BH spins with respect to the orbital angular momentum are then characterized by the polar angles 
\begin{equation}\label{costheta12LB}
\cos\theta_{1}=\hat{\vec{L}}_{\rm B}\cdot\hat{\vec{S}}_1\,,
\qquad
\cos\theta_{2}=\hat{\vec{L}}_{\rm B}\cdot\hat{\vec{S}}_2\,,
\end{equation}
where $\vec{S}_1$ and $\vec{S}_2$ are the spin vectors of the heavier and lighter BH, respectively. 
We can safely assume that the directions $\hat{\vec{S}}_1$ and $\hat{\vec{S}}_2$ remain fixed after the initial draw set by $\vartheta_{\rm max}$, for two reasons:
\begin{enumerate}
\item SN explosions are expected to impart a net linear momentum to newly formed BHs; in the field case, this is the key mechanism that generates spin misalignments by tilting the binary's orbital plane (Sec.~\ref{fieldsec}). However, it appears much less likely that SNe can directly alter the spin direction by exerting a significant torque on the newly formed BH. There is both seminal~\cite{1998Natur.393..139S} and recent~\cite{2022ApJ...938...66T} work in this direction, where torques induced by SN explosions are referred to as ``spin tossing.'' This is a straightforward generalization of our model, which is presented in Appendix~\ref{tossapp}. %

\item After BH formation, the binary is hardened through a series of impulsive dynamical encounters, during which relativistic effects do not have time to alter the spin directions. The role of relativistic precession between binary formation and GW detection is discussed in Sec.~\ref{effectPN}. 
\end{enumerate}

Long-lived triple systems undergoing Kozai--Lidov oscillations require a different treatment (e.g.~\cite{2018ApJ...863....7R}). %

\subsection{Analytical tilt distribution}\label{jointdistr}

The key insight is that the polar angles $\theta_1$ and $\theta_2$ of each binary system are not independent because the two spin directions share common vectors $\vec{L}_{\rm{B}}$ and  $\vec{L}_{\rm{C}}$. 
These spin-spin correlations are the main predictions of our model.

The joint distribution $p(\cos\theta_1, \cos\theta_2)$ can be derived analytically given the model construction described above.   The full derivation of this expression is presented in Appendix~\ref{label:jointcluster}. %
In short, one needs to compute the probability density for the first and second spins to point along directions confined to a spherical cap defined by the polar angle range $0 \leq \theta \leq \vartheta_{\rm max}$. Imposing that these spin directions form angles $\theta_1$ and $\theta_2$ with a common direction uniformly distributed on the sphere, one gets:
\begin{align}
p(\cos\theta_1,\cos\theta_2)
\!=\!\sum_{\ell=0}^{\infty}
\frac{2\ell\!+\!1}{4}\,
\mu_\ell^2(\vartheta_{\text{max}})
P_\ell(\cos\theta_1)
P_\ell(\cos\theta_2)\,,
\label{eq:jointcluster}
\end{align}
where 
\begin{equation}
    \mu_{\ell}(\vartheta_{\text{max}})=\frac{1}{1-\cos\vartheta_{\text{max}}}\int_{\cos\vartheta_{\text{max}}}^{1} P_{\ell}(\cos\theta')\,\dd \cos\theta'
\end{equation} 
and $P_\ell$ are the Legendre polynomials of degree $\ell$.  This expression satisfies the expected behavior in the limiting cases where $\vartheta_{\text{max}}=0$ and $\vartheta_{\text{max}}=\pi$:
\begin{enumerate}
\item
If $\vartheta_{\text{max}}=0$, one has  $\mu_l(0) = P_\ell(1) = 1$ and therefore, using the completeness property of Legendre polynomials, $p(\cos\theta_1, \cos\theta_2) =  \delta(\cos\theta_1 - \cos\theta_2)/2$. 
This condition corresponds to all spins being perfectly aligned within the cluster, and thus having the same polar tilt.%

\item Conversely, if $\vartheta_{\text{max}}=\pi$, all $\mu_\ell$ coefficients vanish except for $\mu_0(\pi)= 1$, which leads to $p(\cos\theta_1, \cos\theta_2) = 1/4$. The distribution becomes isotropic, and the two polar angles are statistically independent. 
\end{enumerate}

\subsection{Results}
\label{clusterresults}

\begin{figure*}[htbp]
    \centering
    
    \includegraphics[width=0.48\textwidth]{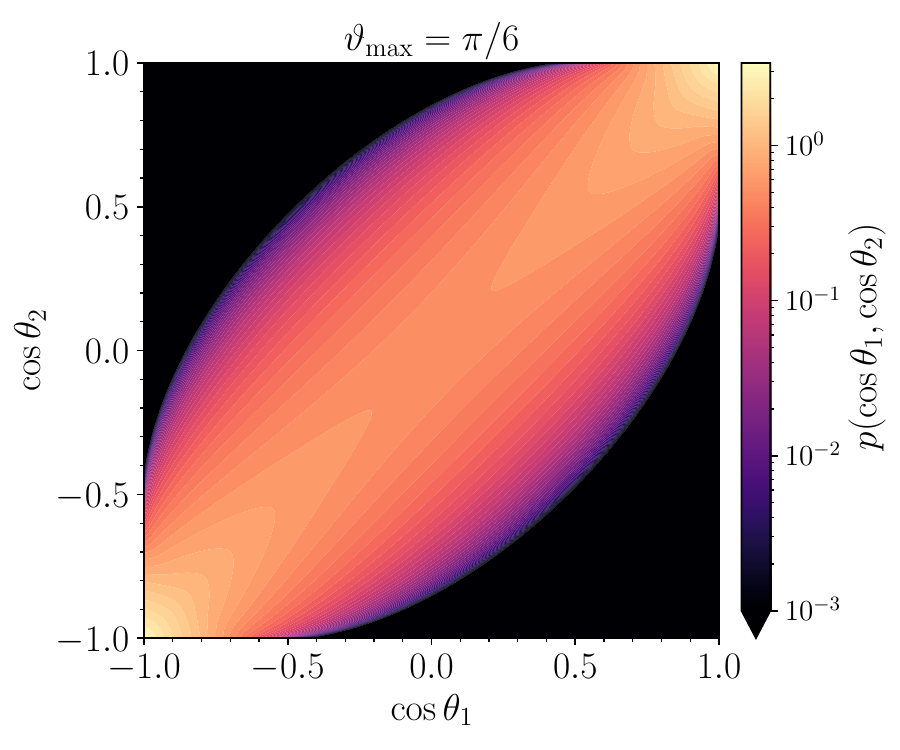}
    \hfill
    \includegraphics[width=0.48\textwidth]{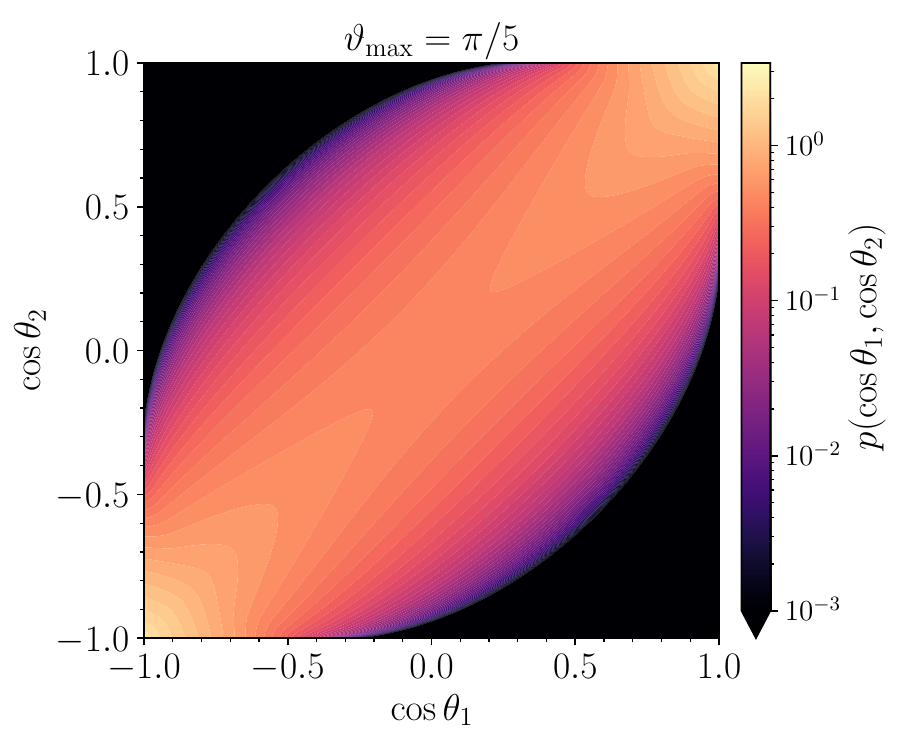}

    \vspace{5pt}
    
    \includegraphics[width=0.48\textwidth]{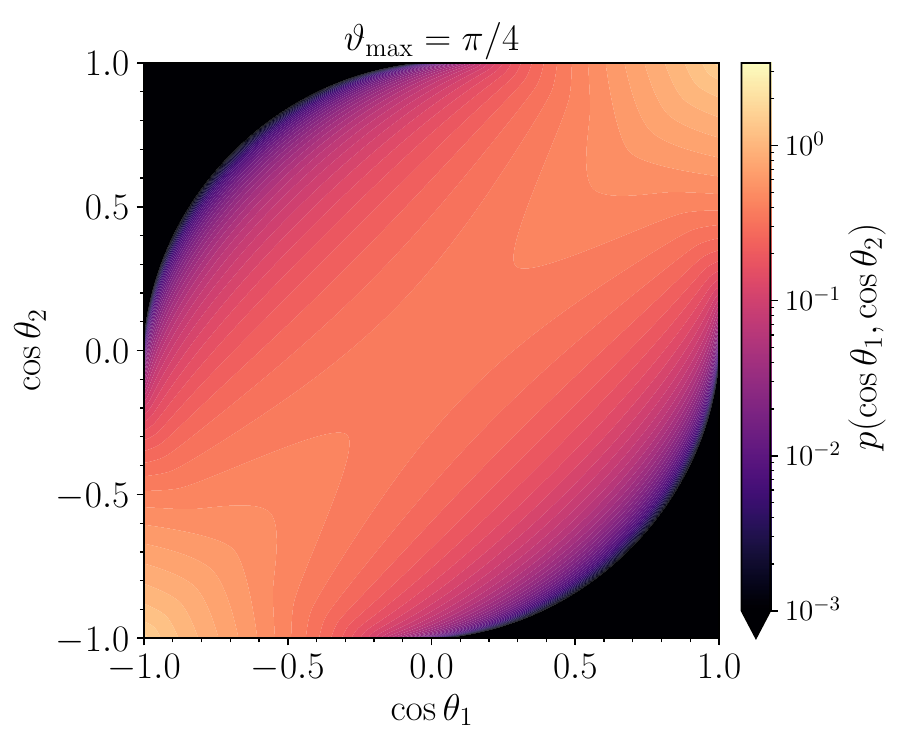}
    \hfill
    \includegraphics[width=0.48\textwidth]{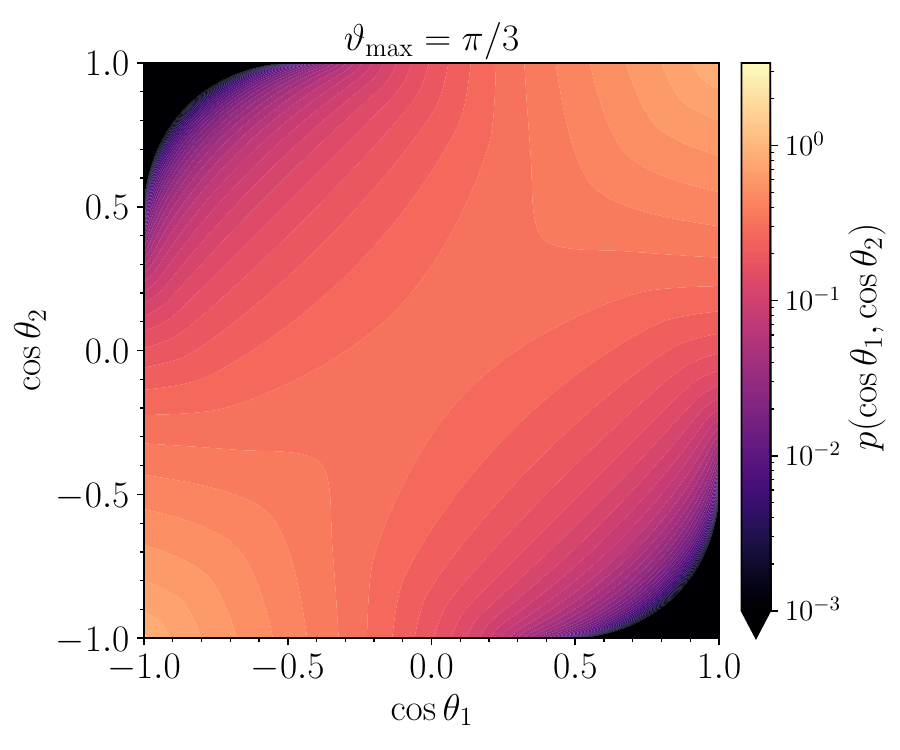}
    
    \vspace{5pt}

    \includegraphics[width=0.48\textwidth]{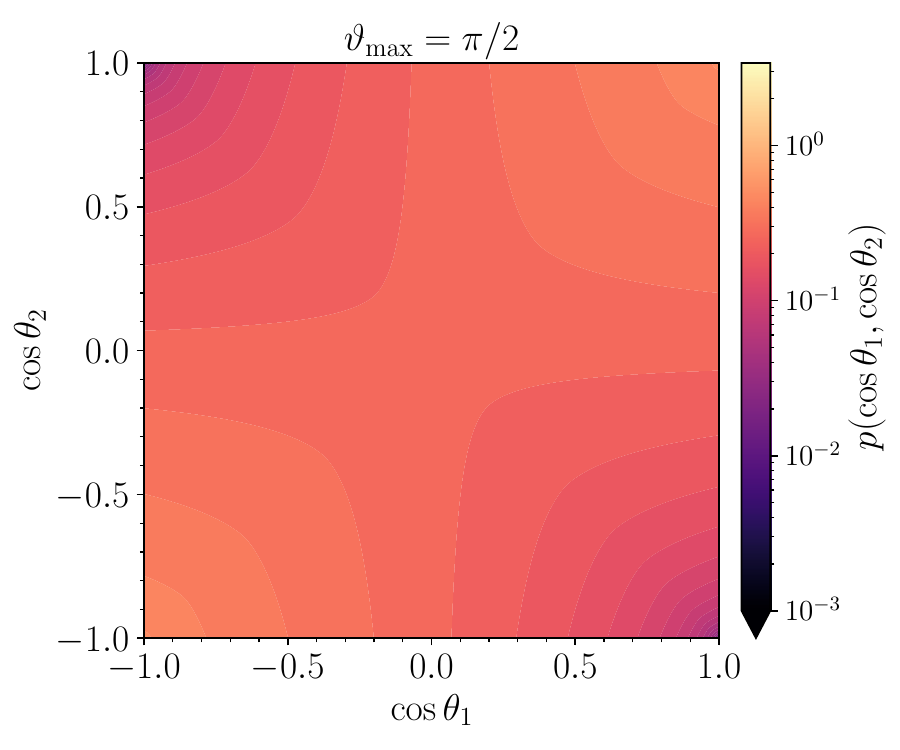}
    \hfill
    \includegraphics[width=0.48\textwidth]{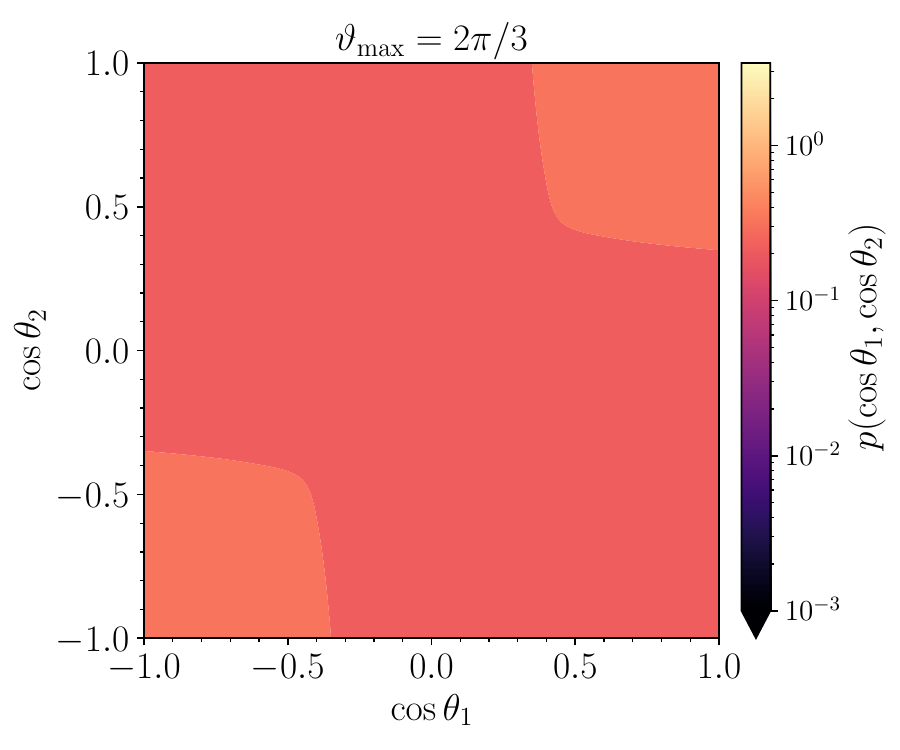}
    
    \caption{Joint distribution $p(\cos\theta_1,\cos\theta_2)$ of the two spin tilt angles for first-generation binary BHs in clusters. In each panel, we vary the angle $\vartheta_{\rm max}$, which controls the degree of alignment between the BH spins and the overall angular momentum of the cluster. %
    }
    \label{fig:jointpdf}
\end{figure*}

Figure \ref{fig:jointpdf} shows $p(\cos\theta_1,\cos\theta_2)$ for different values of $\vartheta_{\rm max}$. For small values of $\vartheta_{\rm max}$, the spins within the cluster are nearly aligned, and the joint distribution is therefore concentrated along the diagonal. As $\vartheta_{\rm max}$ increases, the polar angles become progressively less correlated, causing the joint distribution to broaden. In the limiting case $\vartheta_{\rm max}=\pi$, the distribution becomes uniform over the plane. %

It is important to stress that the marginal distributions of $p(\cos\theta_1) = \int p(\cos\theta_1,\cos\theta_2)  \dd\!\cos\theta_2$ and $p(\cos\theta_2) = \int p(\cos\theta_1,\cos\theta_2)  \dd\!\cos\theta_1$ are, however, uniform for all values of $\vartheta_{\rm max}$. This result can be derived by explicitly integrating the expression above (cf. Appendix~\ref{label:jointcluster}) and follows intuitively from the fact that $\theta_1$ and $\theta_2$ are angles between the spin directions and a randomly oriented orbital angular momentum, see Eq.~(\ref{costheta12LB}). 
Current mainstream population models in GW astronomy capture only such marginal distributions, assuming the two tilts are independently drawn~\cite{2026arXiv260527226T}. We argue that information on cluster initial kinematics and dynamics lies instead in the correlations between the two tilt angles.

\subsection{Hierarchical mergers}
\label{hiermerg}

The model presented above describes BHs whose spins are inherited from those of their stellar progenitors. However, if some BH remnants are retained within the cluster {and dynamically pair up with other BHs}, hierarchical mergers can occur, in which case the argument above no longer applies.

Let us denote the total angular momentum of a BH binary by $\vec{J}=\vec{L}_{\rm B}+\vec{S}_1+\vec{S}_2$. At binary formation, this is largely dominated by the orbital component, $\vec{J}\simeq \vec{L}_{\rm B}$.  During the inspiral, the direction of $\vec{J}$ is conserved to an excellent approximation, with PN calculations showing that, even for fine-tuned resonant configurations, tilts in $\vec{J}$ are of  $\mathcal{O}(10^{-3} \mathrm{rad})$~\cite{2017PhRvD..96b4007Z} (the only possible exception is the even more fine-tuned case of transitional precession, where $|\vec{J}|\simeq 0$~\cite{1994PhRvD..49.6274A}). It follows that, when binaries approach mergers, the ${\vec{J}}$ direction still tracks that of ${\vec{L}}_{\rm B}$ at binary formation, which is isotropically distributed by the many dynamical encounters that lead to binary formation (Sec.~\ref{sec:geometry}). At merger, $\vec{J}$ is largely converted into the spin of the BH remnant, $\vec{S}_{\rm remnant}$, which may then enter the next generation of BH mergers. The spin $\vec{S}_{\rm remnant}$ is therefore also expected to be isotropically distributed. %
 
As a result, the spins of BHs that undergo hierarchical mergers are isotropically distributed on the sphere and no longer retain information about any initial alignment with the cluster angular momentum. In other words, the memory of the initial spin orientation is erased after a single merger generation. For the realistic case of clusters with a finite escape speed, correlations between the spin directions and the imparted BH recoils~\cite{2007PhRvL..98w1102C,2007PhRvL..98w1101G} break perfect isotropy for hierarchical mergers. This point is explored in Sec.~\ref{finite_esc}. 
While our argument applies to clusters, or more broadly to systems that are roughly spherically symmetric, hierarchical BHs formed in axisymmetric environments such as AGN disks can instead approach merger with strongly aligned spins~\cite{2019PhRvL.123r1101Y,2023PhRvD.108h3033S}.

Our prediction for the spin distribution of BH binaries in clusters is therefore a mixture model: the first-generation component follows the distribution presented in Sec.~\ref{jointdistr}, while the hierarchical-merger component is fully isotropic:
\begin{align}
&p_{\rm clusters}(\cos\theta_1,\cos\theta_2 | \vartheta_{\rm max}, \eta_{\rm hm}) 
\notag \\ 
&\qquad=(1-\eta_{\rm hm})p_{\rm 1g}(\cos\theta_1,\cos\theta_2 | \vartheta_{\rm max}) + \frac{\eta_{\rm hm}}{4}\,,
\label{hier_mix}
\end{align}
where $\eta_{\rm hm}$ is the fraction of hierarchical mergers.

\section{Not all spins of BH binaries formed in isolation are aligned}\label{fieldsec}
 
In this section, we present our model for the spin orientations of BH binaries formed in isolation, relaxing the common assumption of perfect alignment. Some steps of the derivation are presented in Appendix~\ref{label:anglefield}.

\subsection{Spin evolution}

\begin{figure*}
  \centering
  \includegraphics[width=0.8\textwidth]{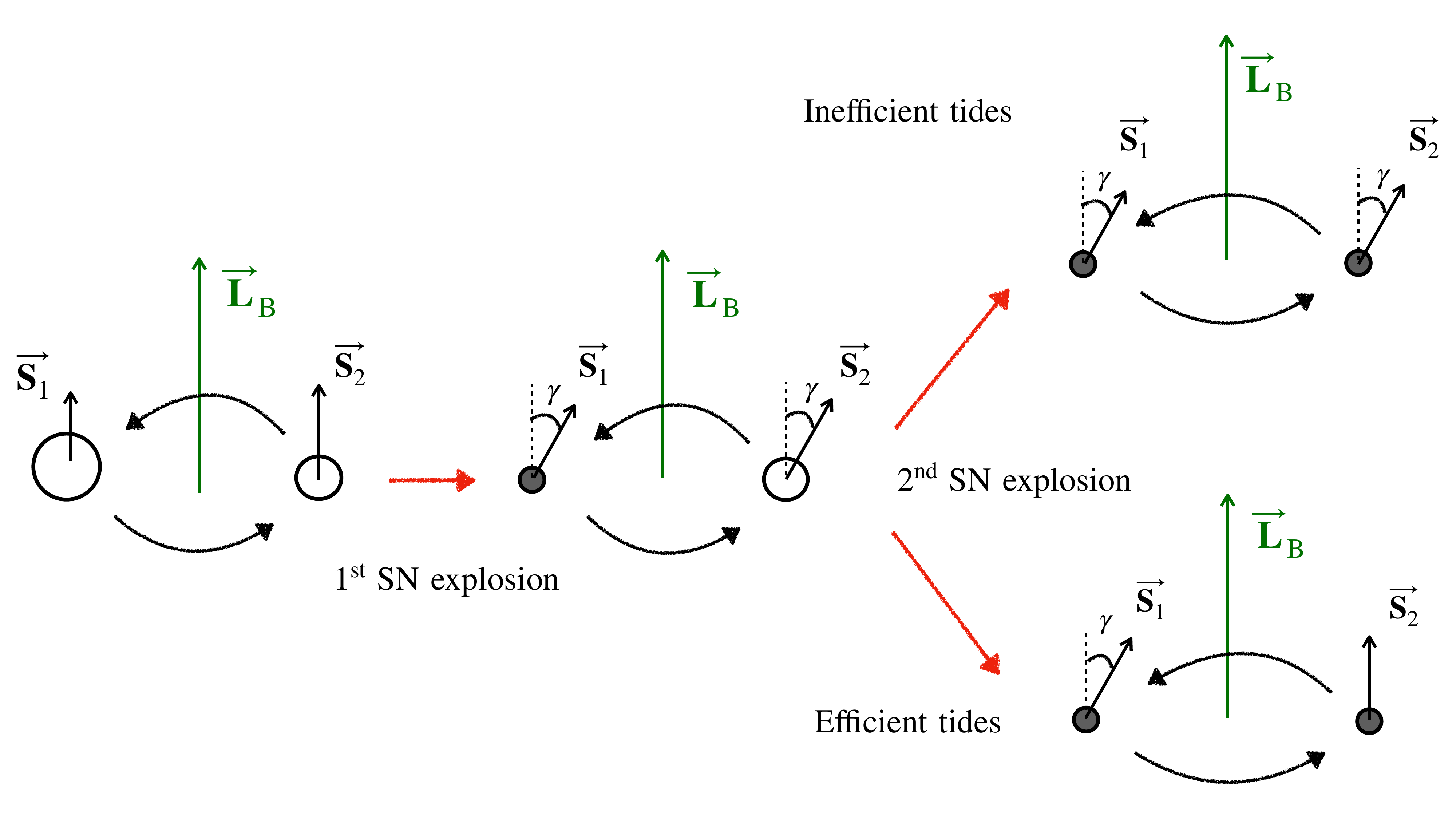}
  \caption{Schematic representation of our model for the spin directions of BH binaries formed in isolation. Two main-sequence stars start with spins aligned with the orbital angular momentum $\vec{L}_{\rm B}$ (left). After the first SN explosion, the resulting tilt of the orbital plane introduces a misalignment angle $\gamma$ for both spins relative to $\vec{L}_{\rm B}$ (center). 
  If tidal interactions are efficient, the secondary's spin $\vec{S}_{2}$ undergoes realignment with $\vec{L}_{\rm B}$ (bottom-right), whereas it remains misaligned otherwise (top-right). The tilt from the second SN explosion is neglected. 
  For simplicity, the 1 and 2 labels in this figure assume no mass ratio reversal.
  }
  \label{fig:fig2}
\end{figure*}

Our model is illustrated in Fig.~\ref{fig:fig2}. Initially, we assume that strong binary interactions, such as tides and mass transfer, have circularized the orbit, and we therefore model the system as two main-sequence stars in a circular binary. The spins of the stars that will form the primary and secondary BHs, $\vec{S}_1$ and $\vec{S}_2$, respectively, are aligned with the initial orbital angular momentum. These do not necessarily correspond to the spins of the initially more and less massive stars, respectively, because mass-ratio reversal may occur during advanced stages of %
binary
 evolution~\cite{2022ApJ...938...45B,2022MNRAS.517.2738M}.

The initially more massive star evolves first and collapses into a BH. During the SN explosion, the newly formed BH receives a natal kick due to asymmetric ejection of matter and neutrinos~(see e.g. Ref.~\cite{Burrows:2020qrp} for a review on the theory of core-collapse SN explosion).
The resulting orbital-plane tilt $\gamma$ corresponds to the spin-orbit misalignment of both binary components and is given by~\cite{2000ApJ...541..319K,2002MNRAS.329..897H,2017PhRvL.119a1101O}
\begin{equation}
    \cos\gamma = \frac{1+u_k\cos\theta_k}{\sqrt{(1+u_k\cos\theta_k)^2+u_k^2\sin^2\theta_k\cos^2\phi_k}}\,.
    \label{eq:cosg}
\end{equation}
Here, $\theta_k$ is the polar angle between the kick velocity $\vec{v}_{\rm k}$ and the orbital velocity $\vec{v}$, the axis $\phi_k$ = 0 is chosen to be parallel to the orbital angular momentum %
and $u_k=\left|\vec{v}_{\rm k}\right|/\left|\vec{v}\right|$ is the ratio between the kick velocity and the orbital velocity.

For simplicity, we draw the kick direction from an isotropic distribution, following the standard assumption in the GW literature. At least for neutron stars, however, it has been suggested that kicks may be preferentially aligned with the stellar spin. Such models appear to reproduce observations of pulsars~\cite{2005MNRAS.364.1397J} and X-ray binaries~\cite{2025arXiv250508857V}, but are in tension with the eccentricities of neutron-star binaries inferred from Gaia observations~\cite{2026ApJ..1002...67B}. For a review of SN kicks, see Ref.~\cite{2025NewAR.10101734P}.
Introducing a maximum opening angle for $\vec{v}_{\rm k}$ around the pre-SN spin $\vec{S}$, in analogy with the parameter $\vartheta_{\rm max}$ of Sec.~\ref{clustersec}, is an interesting generalization of our model.

Following the first SN explosion, the system continues to evolve, potentially undergoing stable and/or unstable mass transfer phases. During these potential stages, tidal interactions can circularize the orbit and align the spin of the extended star with the orbital angular momentum. 
As in Ref.~\cite{2013PhRvD..87j4028G}, we explore both extreme scenarios: complete alignment (“efficient tides” in Fig. \ref{fig:fig2}) and no alignment (“inefficient tides” in Fig. \ref{fig:fig2}). 

When the companion star also completes its evolution and explodes as a SN, a second tilt is imparted to the system. This can further reorient the orbital plane; however, as for the first SN, the resulting tilt depends on the ratio between the kick velocity and the orbital velocity.
By this stage, following possible common-envelope evolution or stable mass transfer, the binary separation has decreased by a factor $\Delta r$. The dimensionless kick velocity $u_k\propto 1/|\vec{v}|\propto \sqrt{r}$ %
therefore decreases by a factor $ \propto \sqrt{\Delta r}$. As a result, the tilt cosine 
$\cos\gamma = 1 - ( u_k \sin\theta_k \cos\phi_k)^2 / 2  + \mathcal{O}(u_k^3)$ %
due to the second SN is factor
$\propto \Delta r$ closer to unity than that of the first.
Population-synthesis calculations show that the binary separation between the two SN events can decrease by $\Delta r\sim 10^3$ (e.g.~\cite{2016Natur.534..512B}), implying that the tilt induced by the post-mass-transfer SN is negligible (see also~\cite{2018PhRvD..98h4036G}). 
We therefore treat the first SN as the dominant source of spin-orbit misalignment in isolated binaries. 

\subsection{Bound orbits}

Since our interest lies in binary systems that merge and produce GW signals, we restrict our analysis to systems that survive the SN explosion.
Indeed, in some cases, the direction and intensity of the imparted kick are such that the post-SN eccentricity exceeds unity and the binary does not remain bound. 
Given $\beta=M_f/M_i\leq 1$ the fraction of the retained mass after the explosion, the condition for keeping the system bound is~\cite{2017PhRvL.119a1101O}
\begin{equation}
    2\beta-1-u_k^2-2u_k\cos\theta_k>0\,. 
    \label{eq:survivecond} 
\end{equation}
Equation (\ref{eq:survivecond}) implies that: 
\begin{enumerate}
    \item If $u_k>1+\sqrt{2 \beta}$ or $-1+\sqrt{2\beta}<0$, the binary is unbound by the SN kick regardless of its direction. 
    \item If $u_k<-1+\sqrt{2\beta}$ and $-1+\sqrt{2\beta}>0$, the binary remains bound regardless of the direction of the SN kick. 
    \item If $-1+\sqrt{2\beta}<u_k<1+\sqrt{2\beta}$ and $-1+\sqrt{2\beta}>0$, a fraction
    \begin{equation}\label{bound}
    f=\frac{1+\cos\theta_{k,\text{max}}}{2} 
\end{equation}
of binaries remain bound. The kick must be outside a cone oriented along the orbital velocity and with angular aperture $\theta_{k,\mathrm{max}}=\arccos[(2\beta-1-u_k^2)/2u_k]$. 
\end{enumerate}

Combining Eqs.~(\ref{eq:cosg}) and (\ref{eq:survivecond}) allows us to infer the tilt magnitude for surviving binaries across different $u_k$ regimes:
\begin{enumerate}
    \item If $u_k \leq 1$, the minimum tilt angle $\gamma_{\rm min}=0$ corresponds to kicks in the orbital plane that do not disrupt the binary and the maximum tilt $\gamma_{\max}=\arccos \small(\sqrt{1-u_k^2} \small) $ corresponds to the case $\phi_k=0$, meaning kicks perpendicular to the orbital plane. 
    \item If $1<u_k\leq\sqrt{1+2\beta}$, the tilt angle $\gamma$ spans the full range from $0$ to $\pi$. In this regime, kicks with $u_k>1$ can produce orbital counter-alignment, while kicks confined to the orbital plane do not disrupt the binary.
    \item If $\sqrt{1+2\beta}<u_k<1+\sqrt{2\beta}$, the kick is sufficiently large that achieving a bound solution implies a non-zero tilt of the orbital plane. The misalignment angle range from 
   $
        \gamma_{\min}=\arccos\!\left[(1/2+\beta-u_k^2/2)/\sqrt{2\beta}\right]
        $
    when $\phi_k=0$ to $\gamma_{\rm max} = \pi$. 
\end{enumerate}
\subsection{Analytical tilt distribution}\label{Field_pdf}
Starting from Eq.~(\ref{eq:cosg}), we analytically derive the probability distribution function of $\cos\gamma$ for (i) a fixed value of  $u_k$, (ii) assuming isotropically distributed SN-kick directions, and (iii) conditioning on binary survival. The full calculation is presented in Appendix~\ref{label:anglefield}; the result is \begin{equation}
\begin{split}
p( \cos\gamma |u_k,\beta)
&=\frac{2(G[\rho_{\rm max}(\cos\gamma )]-G[\rho_{\rm min}(\cos\gamma )])}{\pi \, u_k (1+\cos\theta_{k,\text{max}})\sqrt{1-\cos^2\gamma}}\,,
\label{eq:fieldpdf}
\end{split}
\end{equation}
where
\begin{align}
\label{antiderivative}
\notag G(\rho)&=\cos\gamma \, \arcsin\left[\frac{\rho-\cos\gamma }{\sqrt{u_k^2+\cos^2\gamma -1}}\right]
\\ &-\sqrt{u_k^2 -1- \rho^2 + 2 \rho \cos\gamma}\,, 
\end{align}
while $\rho_{\rm max}$ and $\rho_{\rm min}$ are the maximum and minimum of the interval defined by
\begin{align}
\begin{cases}
\rho\geq 0\,,
\\
|  \rho - \cos\gamma | \leq \sqrt{u_k^2+\cos^2\gamma -1}\,,
\\
1-u_k\leq \rho \cos\gamma \leq 1+u_k\cos\theta_{k,\text{max}}\,.
\end{cases}
\label{intervalrho}
\end{align}

\subsection{Results}
\label{fieldresults}

\begin{figure}
    \includegraphics[width=\columnwidth]{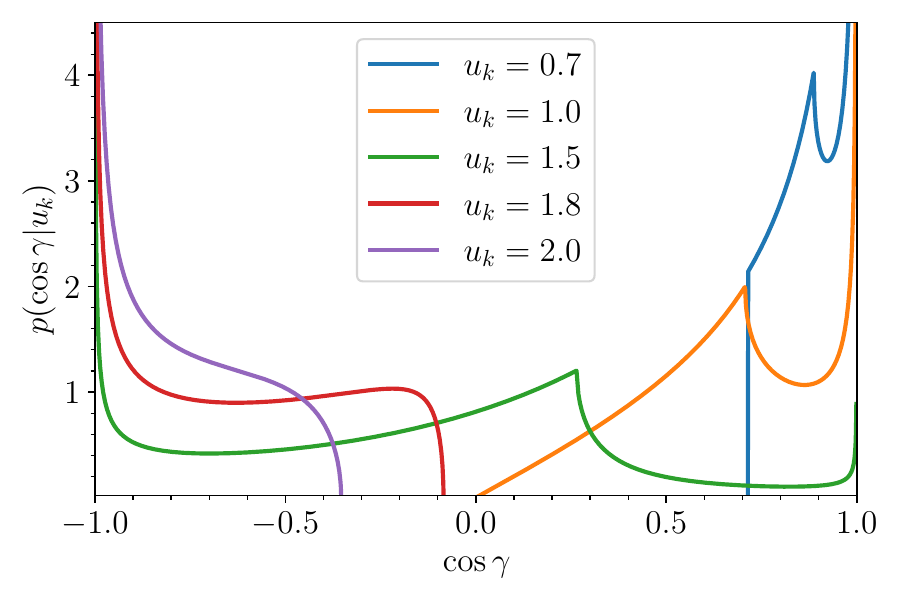}
    \vspace{5pt}
    \includegraphics[width=\columnwidth]{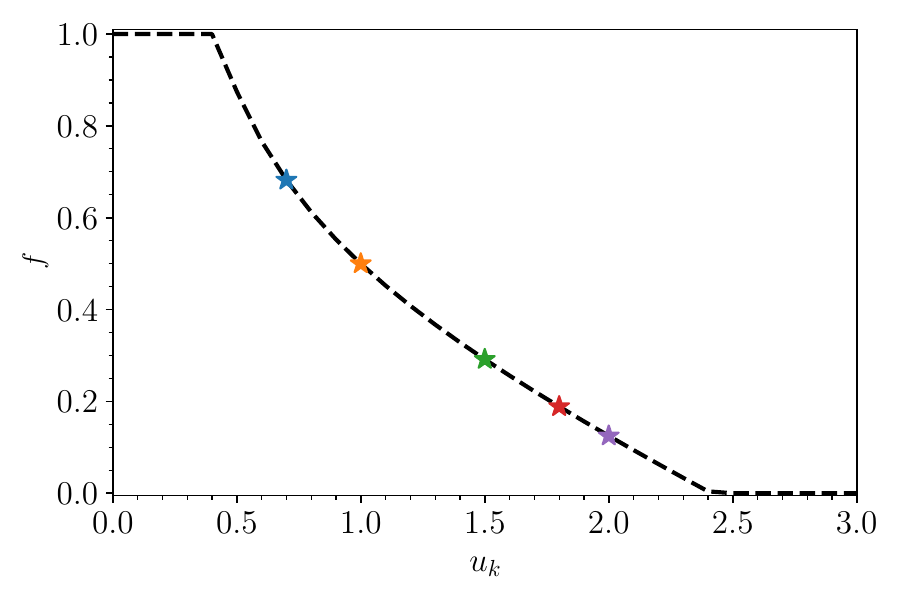}
    \caption{Top panel: distributions of the orbital-plane tilt $p(\cos\gamma|u_k)$ for binaries that survive the SN explosion, shown for different values of $u_k$. In field-binary models where tides are efficient (inefficient), the angle $\gamma$ is equal to both (one of) the BH spin-orbit tilts $\theta_{1,2}$. Bottom panel: fraction of binaries that remain bound as a function of $u_k$. Star markers indicate the values of $u_k$ corresponding to the distributions shown in the top panel.}
    \label{fig:field}
\end{figure}

The top panel of Fig.~\ref{fig:field} shows the distribution of the SN-induced orbital-plane tilt $p(\cos\gamma | u_k)$ for binaries that survive the SN explosion, for different values of $u_k$ and assuming $\beta=1$. For small values of $u_k$, the tilt imparted to the orbital plane is modest, and the distribution is therefore confined to a relatively narrow range. As $u_k$ increases, the distribution broadens. For larger values of $u_k$, only configurations in which the orbital angular momentum is nearly reversed with respect to its pre-SN direction remain bound. 

Although each distribution is individually normalized to unity, their relative astrophysical importance differs. The fraction of binaries that remain bound is much larger for small values of $u_k$, whereas stronger kicks are increasingly likely to disrupt the system, leaving only a small fraction of surviving systems. The bottom panel of Fig.~\ref{fig:field} shows the fraction of bound binaries $f$ as a function of $u_k$, with stars marking the values of $u_k$ adopted in the top panel. This quantity should be treated as a weight for the corresponding tilt distributions.%

Within this simplified model, BH binaries reach the GW-driven regime with polar spin-orbit angles $\theta_1=\theta_2=\gamma$ when tides are neglected. If instead tides are efficient, the angle of the second-born BH is reset to zero, while that of the first-born BH remains equal to $\gamma$. While the first- and second-born BHs correspond to the heavier and lighter zero-age main-sequence stars, respectively, they do not necessarily correspond to the heavier and lighter BHs if stellar-evolutionary processes reverse the mass ratio during the binary evolution \cite{2022ApJ...938...45B,2022MNRAS.517.2738M}. As for the azimuthal spin angles, while they are identical immediately after the SN, they are rapidly randomized by Newtonian precession on a timescale much shorter than the binary lifetime. Relativistic precession during the late inspiral will, in general, also modify the polar angles $\theta_1$ and $\theta_2$; this effect is explored in Sec.~\ref{effectPN}.

Our model for the spin directions of isolated binaries can be summarized as follows. Referring to the probability distribution of Eq.~(\ref{eq:fieldpdf}) as $p_{\rm tilt}$, one can construct a mixture model to account for tidal efficiency and the occurrence of mass-ratio reversal. This results in the tilt-angle distribution
\begin{align}
&p_{\rm field}(\cos\theta_1,\cos\theta_2 | u_k, \beta, \eta_{\rm tides}, \eta_{\rm rev}) \notag \\&\quad= (1-\eta_{\rm tides}) p_{\rm tilt}(\cos\theta_1 | u_k,\beta) \,\delta(\cos\theta_2-\cos\theta_1)
\notag \\&\quad+  \eta_{\rm tides}(1-\eta_{\rm rev}) p_{\rm tilt}(\cos\theta_1 | u_k,\beta) \delta(\cos\theta_2-1) 
\notag \\&\quad+  \eta_{\rm tides} \eta_{\rm rev}   \delta(\cos\theta_1-1) p_{\rm tilt}(\cos\theta_2 | u_k,\beta) 
\label{pfieldall}
\end{align}
where $\eta_{\rm tides}$ is the fraction of binaries with efficient tides and $\eta_{\rm rev}$ is the fraction of binaries undergoing mass-ratio reversal.

\section{Numerical investigation}
In this section, we test the robustness of our analytical model with some numerical experiments. Our implementation builds on and extends the structure of the \textsc{qluster} code~\cite{2019PhRvD.100d1301G,2023grav.conf...49G}. We apply the same approach to both formation scenarios, except for the treatment of spin directions and the inclusion of hierarchical mergers in the dynamical channel.

\subsection{Numerical model}\label{num_model}

We initialize a collection of $N$ BHs, with masses $m\in\left[m_{\rm min},m_{\rm max}\right]$ drawn from an initial mass function $p(m)\propto m^{\bar\gamma}$ and dimensionless spin magnitudes $\chi\in\left[\chi_{\rm min},\chi_{\rm max}\right]$ drawn from a uniform distribution. 
Mergers are performed by drawing two BHs with masses $m_1 > m_2$ from the collection according to the pairing probability%
\begin{align}
p_{\text{pair}}(m_1) &\propto
\begin{cases}
m_1^{\bar \alpha_1} & \text{if } m_1 < m_{\text{break}}\,, \\
m_1^{\bar \alpha_2} & \text{if } m_1 > m_{\text{break}}\,,
\end{cases}
\\
p_{\mathrm{pair}}(m_2 \mid m_1) &\propto
\left(\frac{m_2}{m_1}\right)^{\bar\beta}\,.
\end{align}
This prescription is loosely inspired by current GW population fits
\cite{2026arXiv260527226T}. Based on those results, we set
$\bar\alpha_1=-1.5$, $\bar\alpha_2=-5.4$, $\bar\beta=1.0$,
$m_{\rm min}=5\,M_\odot$, $m_{\rm max}=50\,M_\odot$,
$m_{\text{break}}=37.5\,M_\odot$, $\chi_{\rm min}=0$,
$\chi_{\rm max}=0.5$, and $\bar\gamma=-2.35$, with the latter
corresponding to a Salpeter initial mass function. %

Once two BHs are paired, the spin orientations are computed as follows:  
\begin{enumerate}
    \item For the dynamical scenario, the model is parametrized by $\vartheta_{\rm max}$. The BH tilt angles $\theta_{1,2}$ are set by the prescription described in Sec.~\ref{sec:geometry}. %
    \item For isolated binaries, the model is parametrized by $u_{\rm k}$. We set $\beta=1$ for simplicity. The BH tilt angles $\theta_{1,2}$ are initially equal to each other and are computed from Eq.~(\ref{eq:cosg}). The effect of tidal interactions is also explored:
    \begin{enumerate}
        \item If tides are efficient, the primary spin retains the SN-induced tilt while the secondary spin is aligned with the orbital angular momentum and so $\theta_1=\gamma$, $\theta_2=0$. %
        \item If tides are inefficient, both spins retain the SN-induced tilt, resulting in $\theta_1=\theta_2=\gamma$. %
    \end{enumerate}
\end{enumerate}
In the plots below, for case (ii.a) we assume that the mass ratio is not reversed during binary evolution; in the case of mass-ratio reversal, results are analogous but $\theta_1\leftrightarrow \theta_2$.
In all cases, the azimuthal angle $\Delta\Phi$ between the projections of the two spins onto the orbital plane is assumed to be uniformly distributed. 

Relativistic spin-spin and spin-orbit couplings during the binary inspiral, from formation to merger, cause the spin orientation to depart from the distributions presented above. Comparing against a numerical procedure that includes these effects is therefore important for testing the robustness of our analytic expressions.
Specifically, we evolve the spin-orientation angles $\theta_1$, $\theta_2$, and $\Delta\Phi$ using the precession-averaged %
 PN equations of motion implemented in the \textsc{precession} code~\cite{2023PhRvD.108b4042G,2016PhRvD..93l4066G}. We evolve the binaries from an initial separation of $r=10^5 G(m_1+m_2)/c^2$ (a rough estimate of the orbital separation at BH-binary formation) to a final separation of $r=10 G(m_1+m_2)/c^2$ (where the PN approximation is expected to break down). 
We verified that our results are not affected by these choices. For simplicity, we assume quasi-circular sources throughout; the interplay between eccentricity and spin directions in PN evolutions has been addressed in Refs.~\cite{2023PhRvD.108l4055F,2024PhRvD.110f3012F,2025PhRvD.112d4067P,2025arXiv250507238S}.

For the dynamical scenario, the contribution of hierarchical mergers is accounted for by assigning a global property, namely the escape speed $v_{\text{esc}}$ of the cluster. For each merger, we estimate the mass, spin magnitude, and kick velocity of the remnant using fitting formulae calibrated to numerical-relativity simulations~\cite{2016PhRvD..93l4066G}. The spin direction of the remnant BH is assumed to be isotropically distributed, as discussed in Sec.~\ref{hiermerg}.
The kick velocity $v_{\text{kick}}$ determines whether the remnant is retained in the cluster: if $v_{\text{kick}}<v_{\text{esc}}$, the BH is placed back into the collection and is available for further mergers; if $v_{\text{kick}}>v_{\text{esc}}$, the BH is removed. This process is iterated until fewer than two BHs remain. Each cluster therefore produces between $N \divisionsymbol 2$ and $N-1$ mergers. For context, the escape speed of typical globular clusters (nuclear star clusters) is $v_{\text{esc}}\lesssim 100$ km/s ($v_{\text{esc}}\lesssim 1000$ km/s)~\cite{2002ApJ...568L..23G,Lutzgendorf:2012tn,Antonini:2016gqe}.
BH mergers are labeled according to the generations of the two BHs involved: mergers between two first-generation BHs are denoted as ``1g+1g,'' mergers between a first- and a second-generation BH as ``1g+2g,'' mergers between two second-generation BHs as ``2g+2g,'' and mergers involving at least one BH of a higher generation as ``$>2$g.''

In the dynamical case, we carry out simulations with $N=5000$ BHs (which is in the converged regime identified in Refs.~\cite{2019PhRvD.100d1301G,2023grav.conf...49G}), averaging results over 200 realizations. %
In the isolated scenario, we %
run sets of $N=10^5$ binaries. %

\begin{figure*}[htbp]
    \centering
    \includegraphics[width=0.33\textwidth]{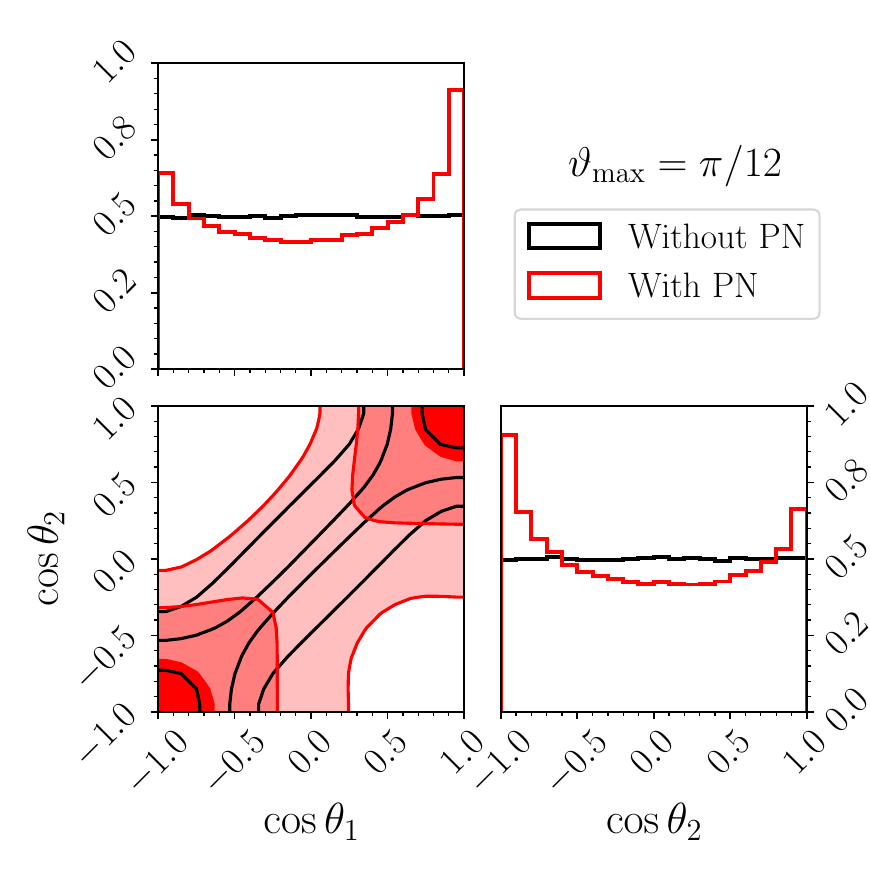}\hspace{1pt}%
    \includegraphics[width=0.33\textwidth]{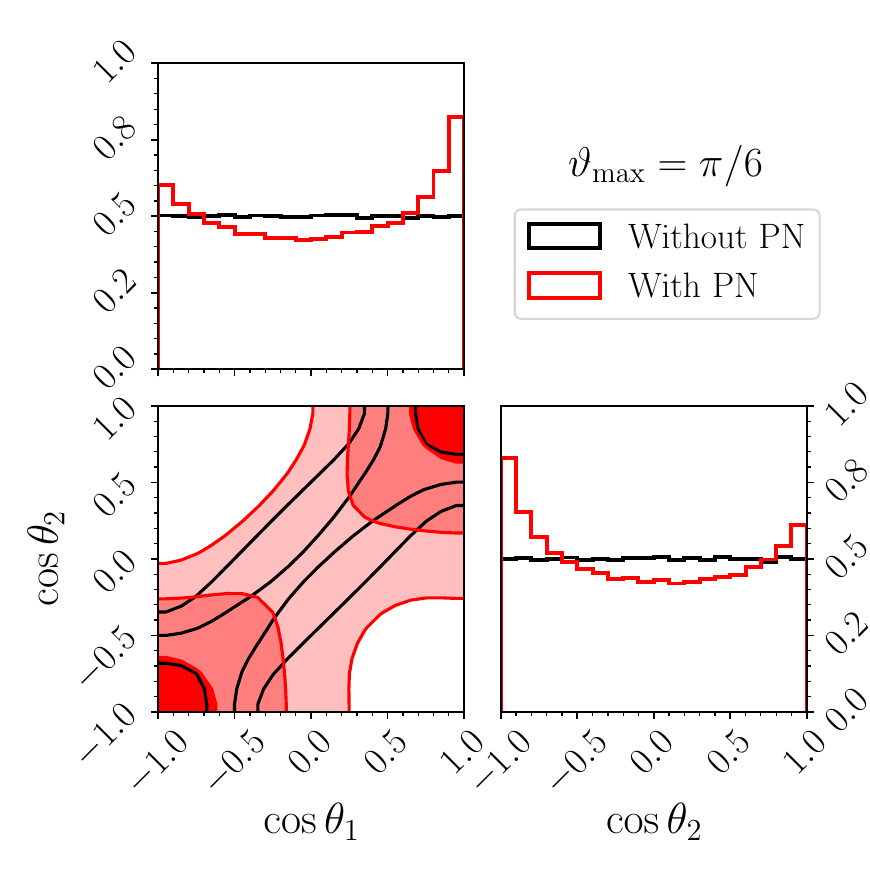}\hspace{1pt}%
    \includegraphics[width=0.33\textwidth]{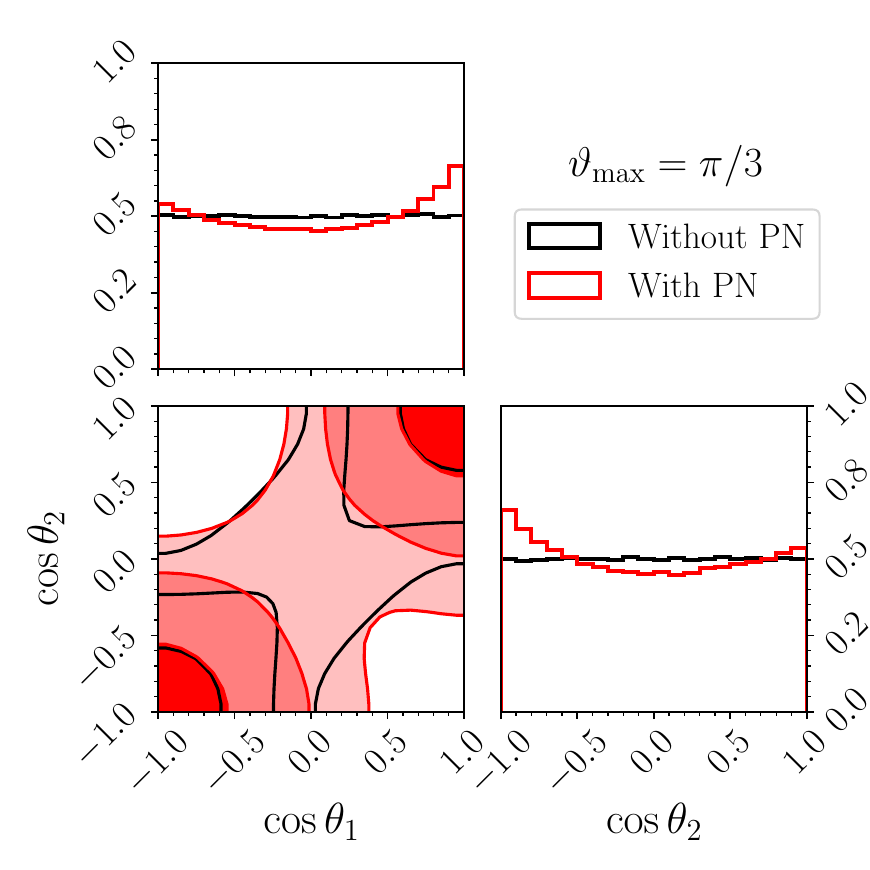}
    \vspace{5pt}
    \includegraphics[width=0.33\textwidth]{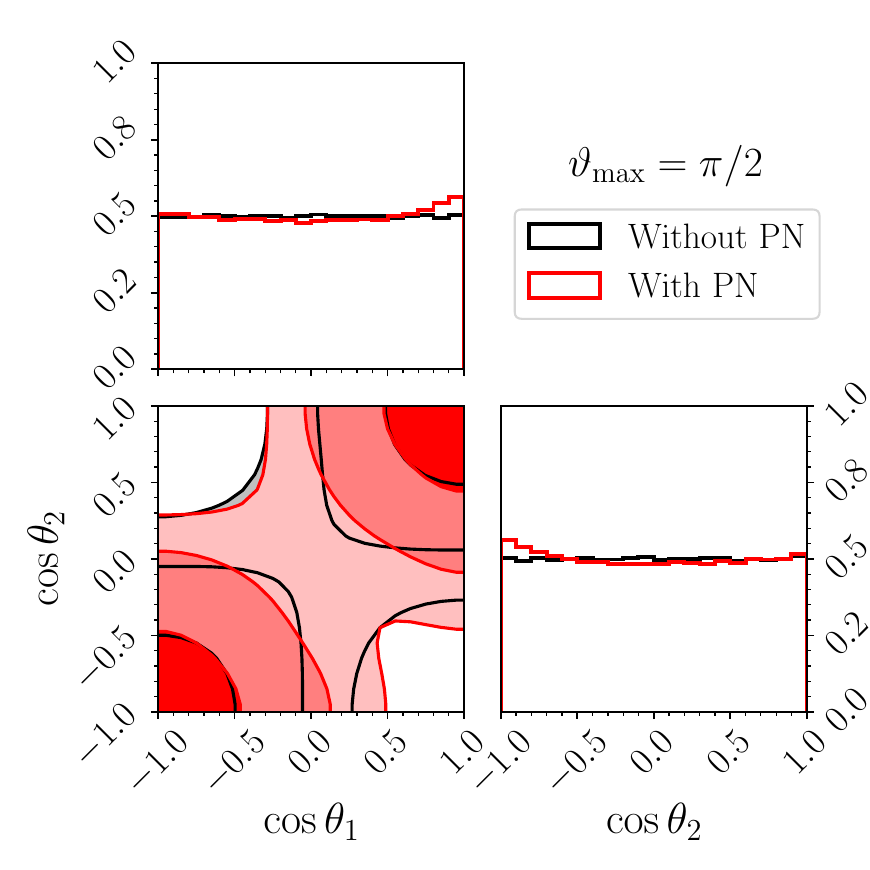}\hspace{1pt}%
    \includegraphics[width=0.33\textwidth]{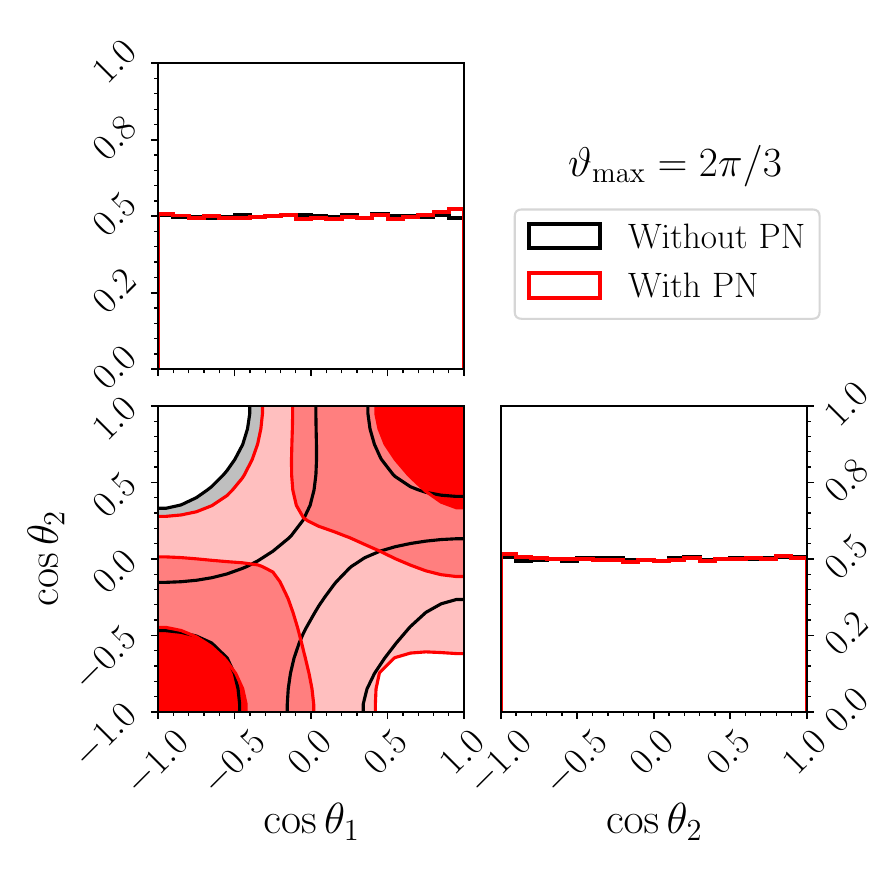}\hspace{1pt}%
    \includegraphics[width=0.33\textwidth]{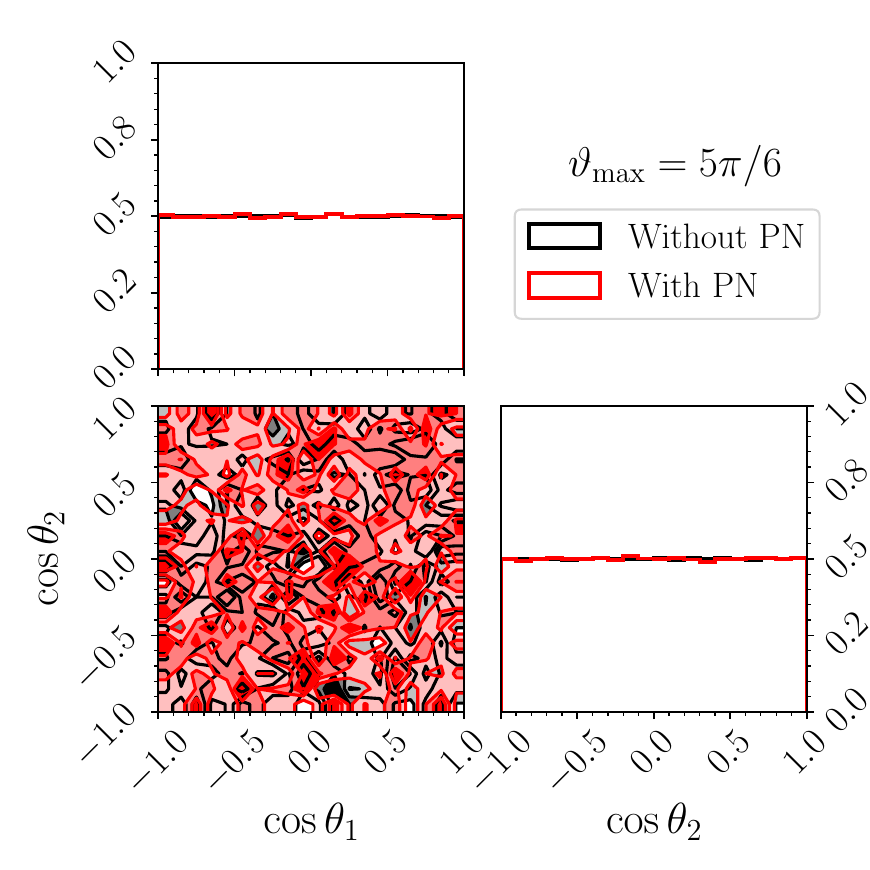}
    \caption{Joint distribution of $\cos\theta_1$ and $\cos\theta_2$ for 1g+1g binaries in the dynamical scenario, shown for increasing values of $\vartheta_{\rm max}$. Contours refer to the regions containing $15\%$, $50\%$, and $85\%$ of the probability. Red (black) curves refer to distributions when the PN evolution between BH binary formation and merger is considered (neglected).}
    \label{fig:ClusterPN}
\end{figure*}

\subsection{Impact of relativistic spin couplings} %
 \label{effectPN}

We first test the impact of PN evolution on the spin orientations in both the cluster and field scenarios. In Figs.~\ref{fig:ClusterPN}, \ref{fig:fieldPN}, and \ref{fig:fieldPN_tides} below, the cases ``without PN'' correspond to the analytical prescriptions derived in Sec.~\ref{clustersec} and \ref{fieldsec} ---which we verified numerically when the PN evolution is swtiched off--- while the cases ``with PN'' are obtained from the numerical setup described in Sec.~\ref{num_model}.

Figure \ref{fig:ClusterPN} considers 1g+1g binaries in the dynamical scenario for different values of $\vartheta_{\rm max}$. Relativistic precession results in broader distributions at merger compared to those at binary formation. While this effect is more pronounced for small values of $\vartheta_{\rm max}$, the variations remain modest overall. This implies that the analytical distribution $p(\cos\theta_1,\cos\theta_2)$ derived in Section \ref{sec:geometry} still provides a good representation of the correlation between $\cos\theta_1$ and $\cos\theta_2$, even when relativistic evolution is taken into account.

Figure \ref{fig:fieldPN} illustrates the effects of PN evolution in the isolated scenario with inefficient tides for different values of $u_k$. Initially, both $\cos\theta_1$ and $\cos\theta_2$ are distributed according to Eq.~(\ref{eq:fieldpdf}), while their joint distribution is confined to the diagonal. Relativistic precession broadens the distribution, which, however, remains concentrated along the $\theta_1=\theta_2$ diagonal.
 
\begin{figure*}[htbp]
    \centering
    \includegraphics[width=0.33\textwidth]{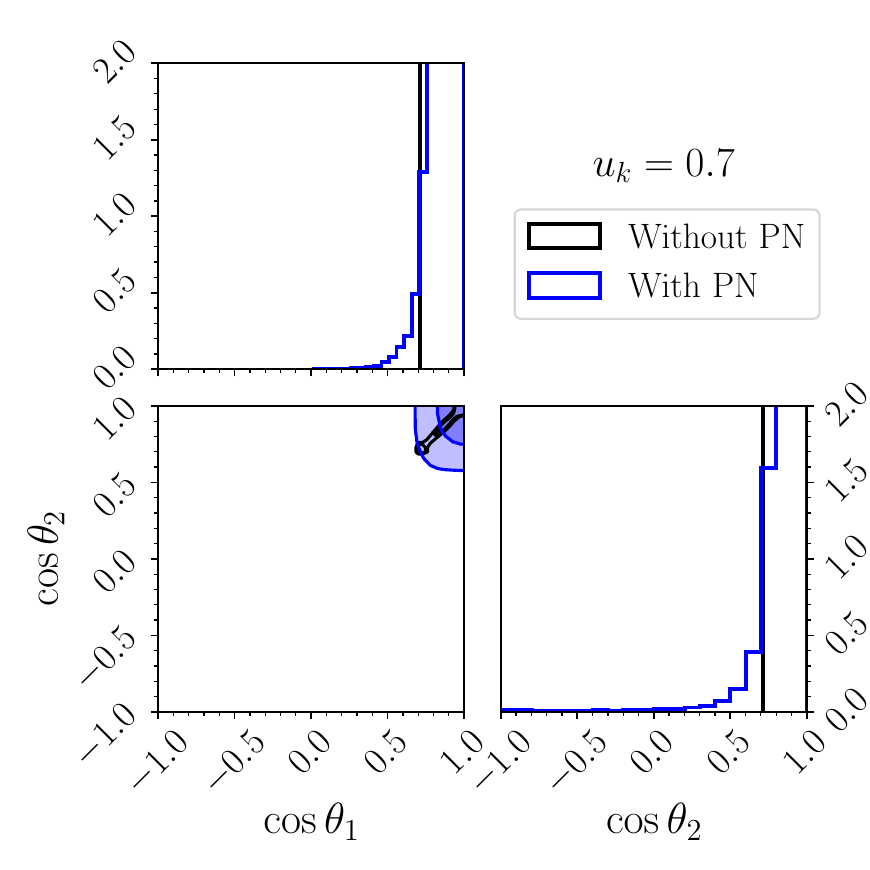}\hspace{1pt}%
    \includegraphics[width=0.33\textwidth]{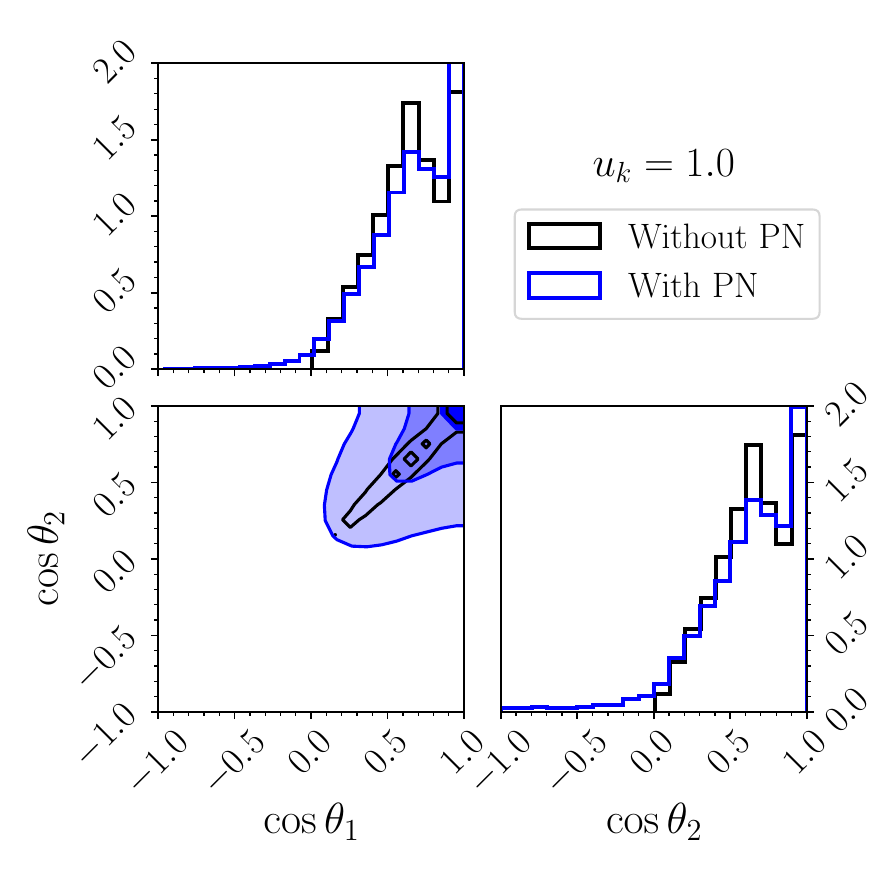}\hspace{1pt}%
    \includegraphics[width=0.33\textwidth]{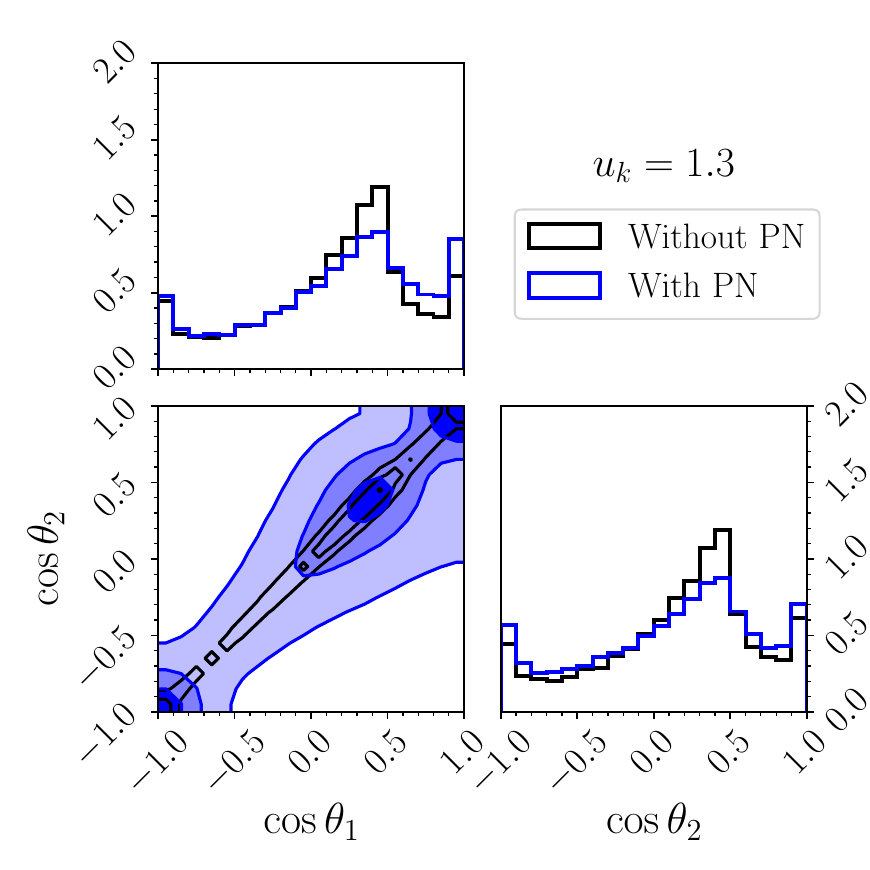}
    \vspace{5pt}
    \includegraphics[width=0.33\textwidth]{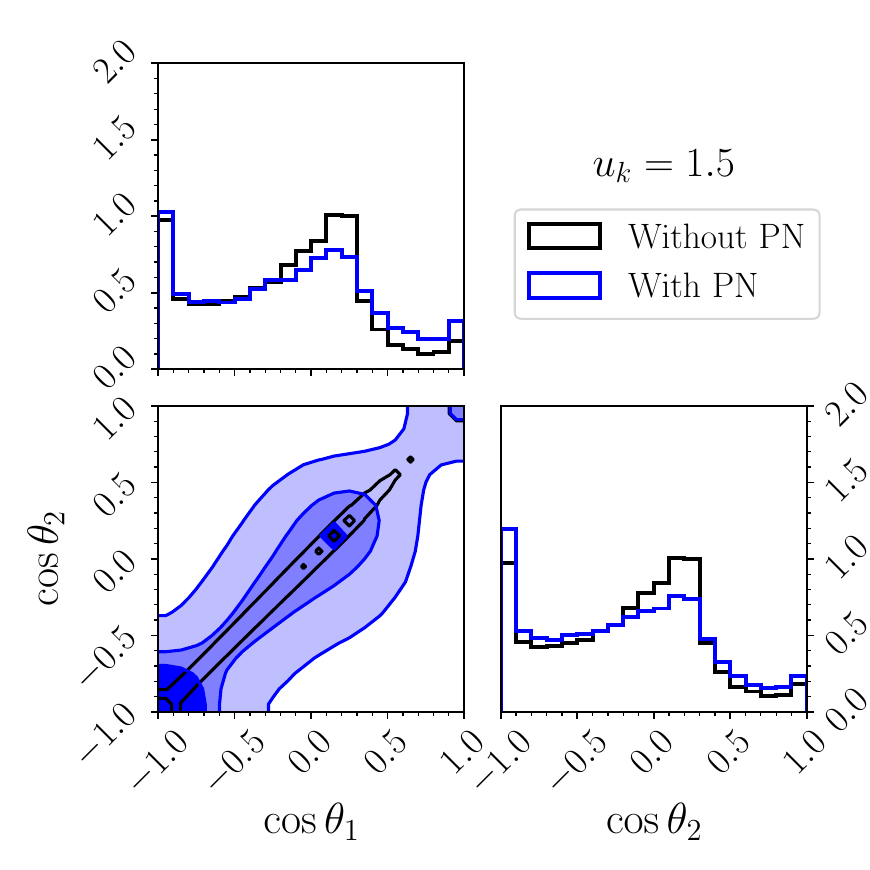}\hspace{1pt}%
    \includegraphics[width=0.33\textwidth]{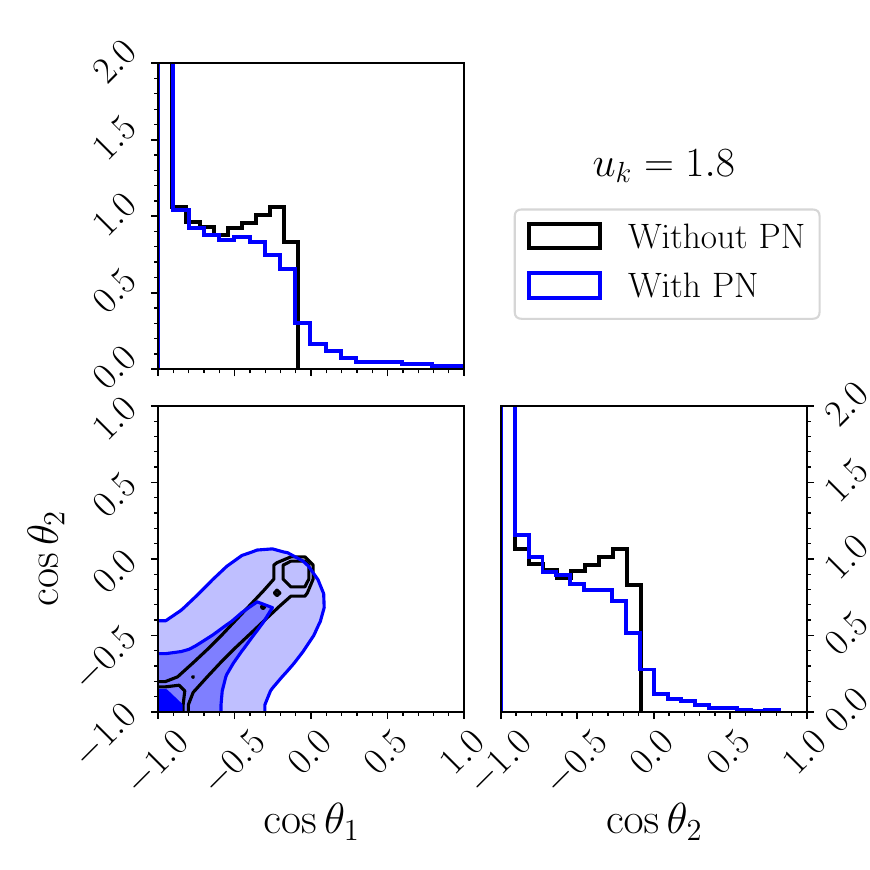}\hspace{1pt}%
    \includegraphics[width=0.33\textwidth]{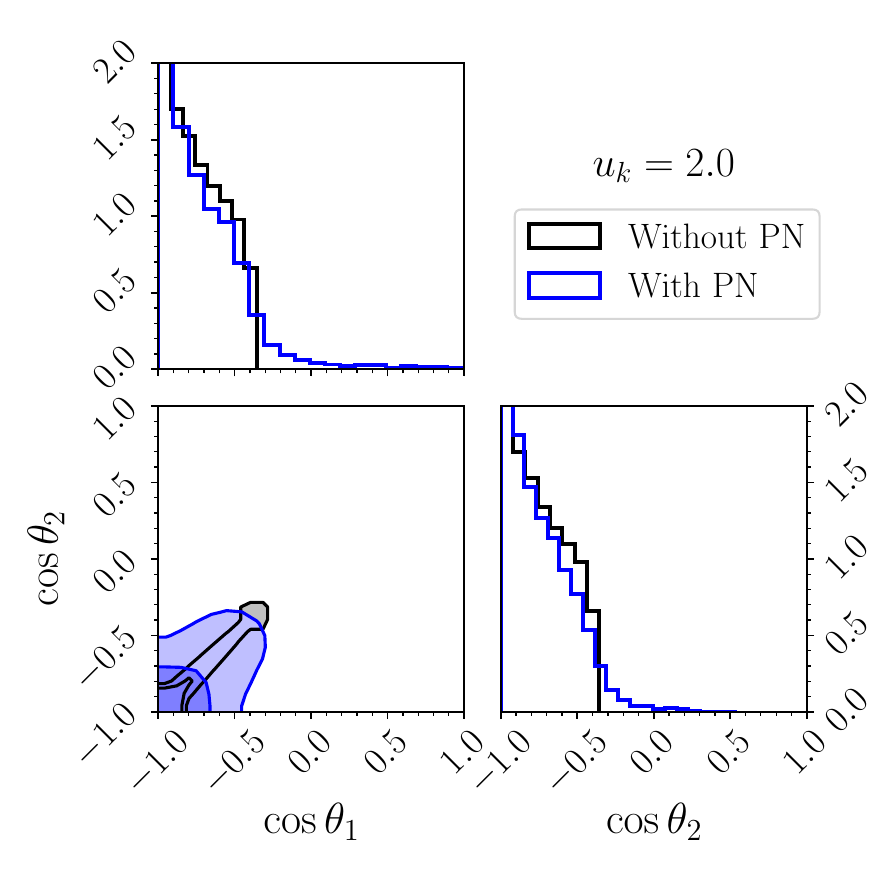}
    \caption{Joint distribution of $\cos\theta_1$ and $\cos\theta_2$ in the field scenario for inefficient tides, shown for increasing values of $u_k$. Contours refer to the regions containing  $15\%$, $50\%$, and $85\%$ of the probability. Blue (black) curves refer to distributions when the PN evolution between BH binary formation and merger is considered (neglected).}    
    \label{fig:fieldPN}
\end{figure*}

Figure \ref{fig:fieldPN_tides} instead shows the effect of relativistic precession in the isolated scenario with efficient tides, assuming the mass ratio is not reversed during binary evolution. In this case, at the beginning of the inspiral, only $\cos\theta_1$ follows Eq.~(\ref{eq:fieldpdf}), whereas $\cos\theta_2=1$. After PN evolution, the distributions of both $\cos\theta_1$ and $\cos\theta_2$ broaden. Compared to the case with inefficient tides, the joint distribution is highly asymmetric.

For the isolated scenario, we therefore find that relativistic precession does not significantly alter the marginal distributions of the two polar angles. By contrast, their joint distribution is more strongly affected by PN dynamics. A possible extension of our model would be to add a small 2D Gaussian perturbation on top of $\cos\gamma$ from Eq.~(\ref{eq:cosg}), which would capture the effect of relativistic spin evolution in a phenomenological fashion, without the need to perform PN integrations. We plan to explore this in future work.

\begin{figure*}[htbp]
    \centering
    \includegraphics[width=0.33\textwidth]{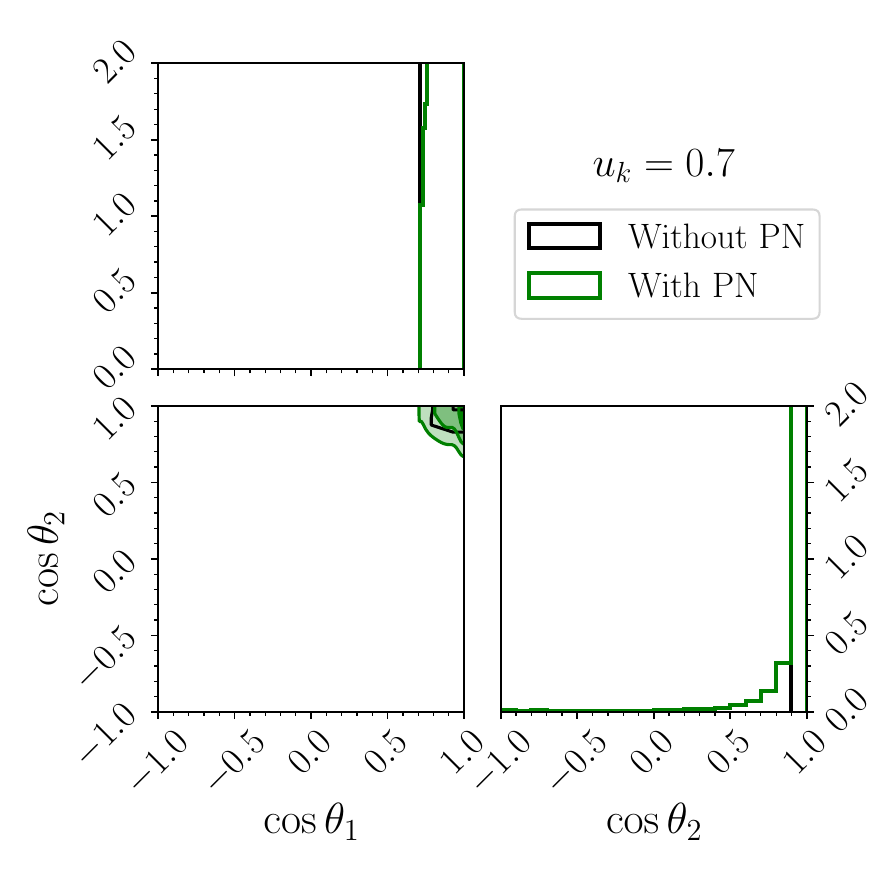}\hspace{1pt}%
    \includegraphics[width=0.33\textwidth]{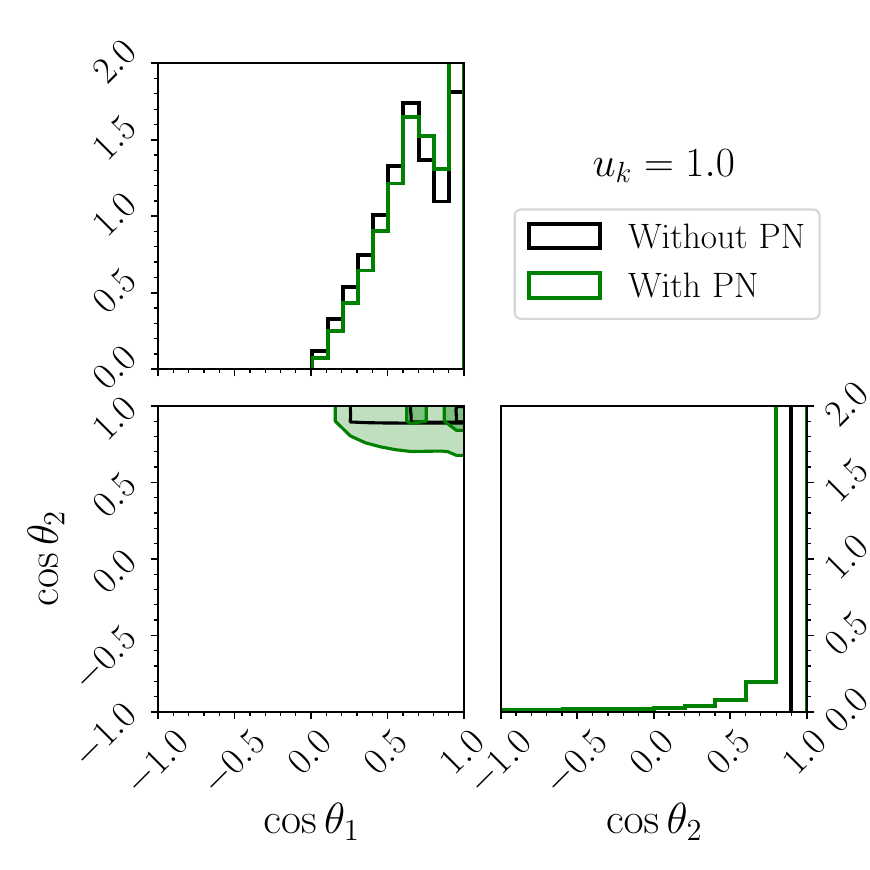}\hspace{1pt}%
    \includegraphics[width=0.33\textwidth]{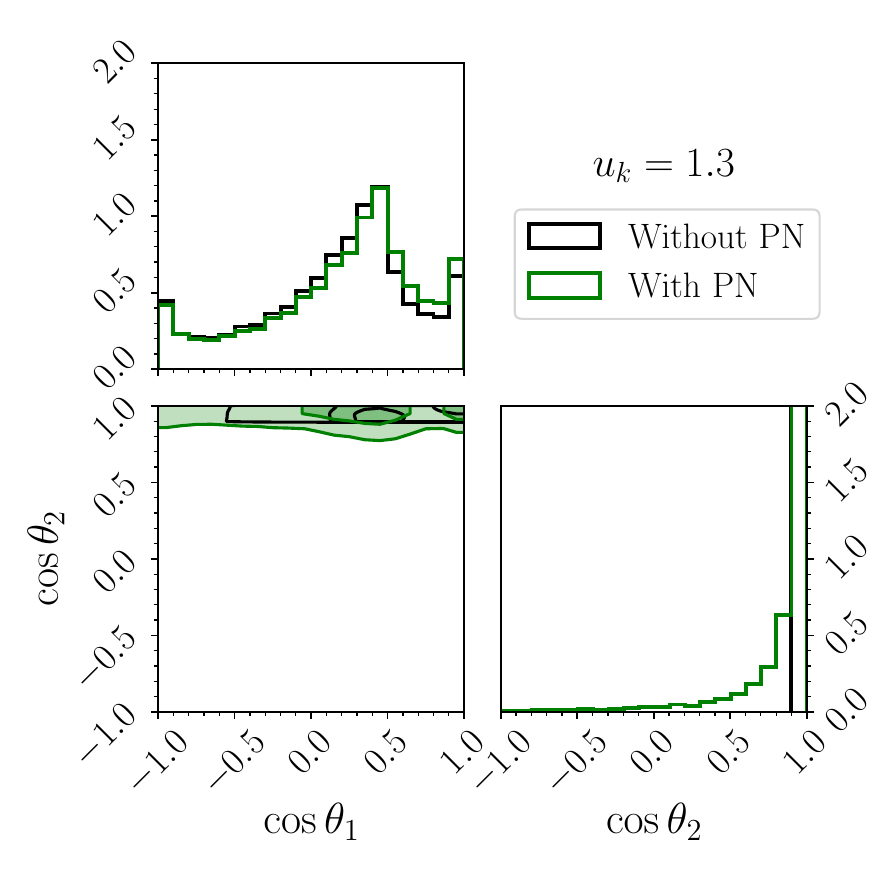}
    \vspace{5pt}
    \includegraphics[width=0.33\textwidth]{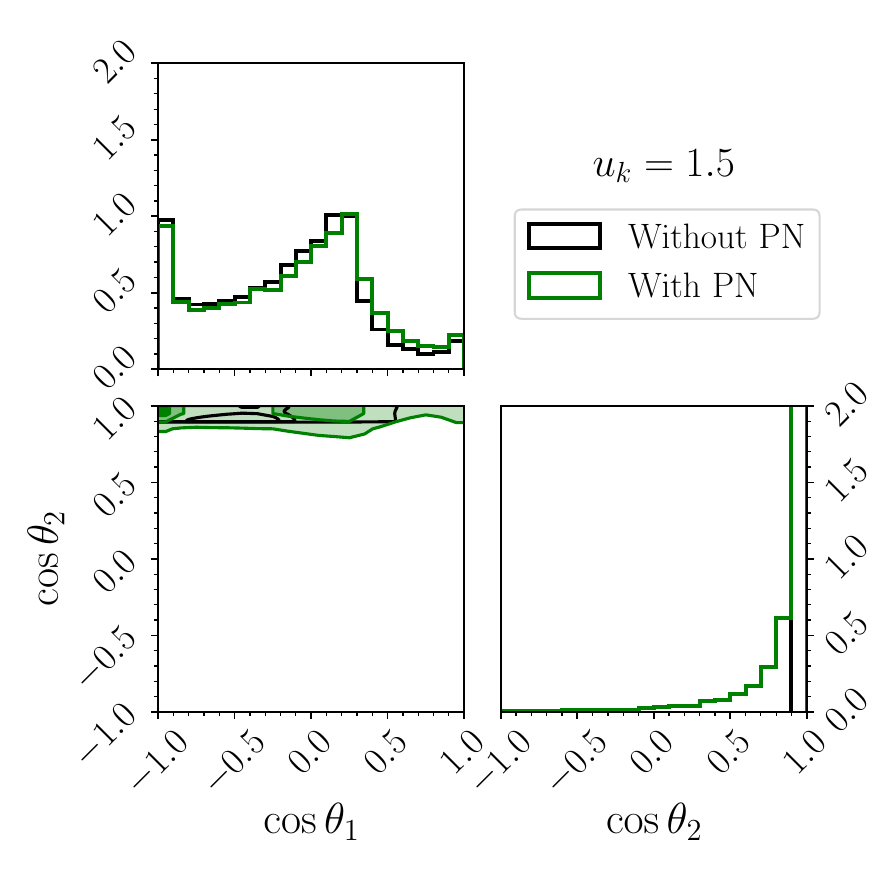}\hspace{1pt}%
    \includegraphics[width=0.33\textwidth]{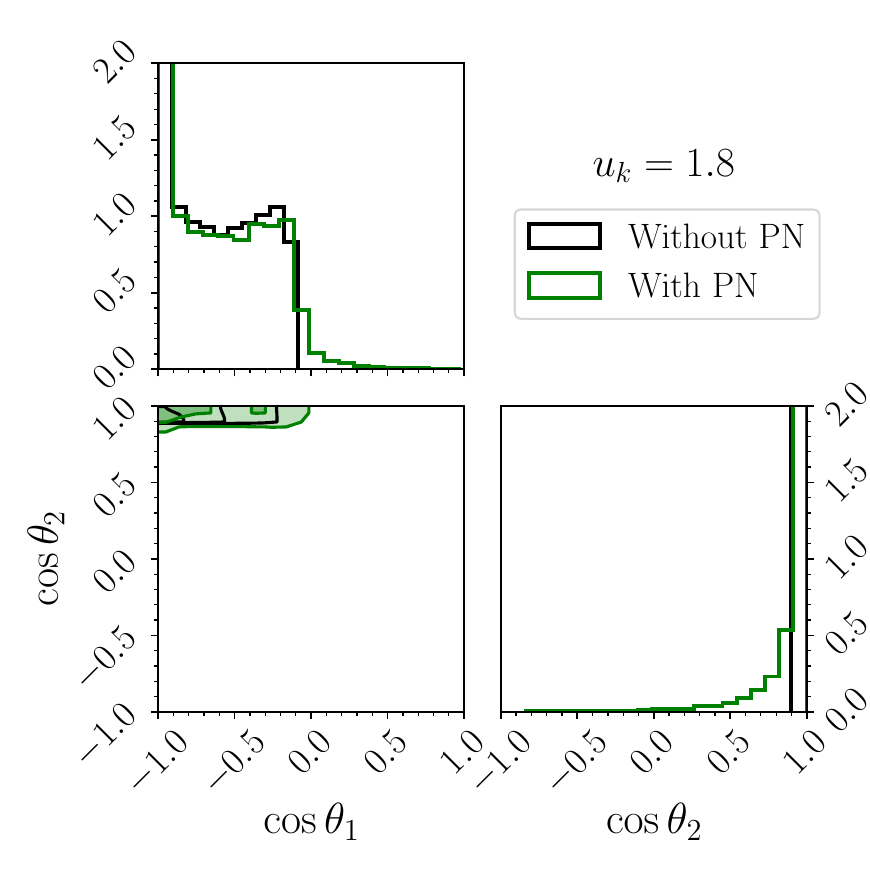}\hspace{1pt}%
    \includegraphics[width=0.33\textwidth]{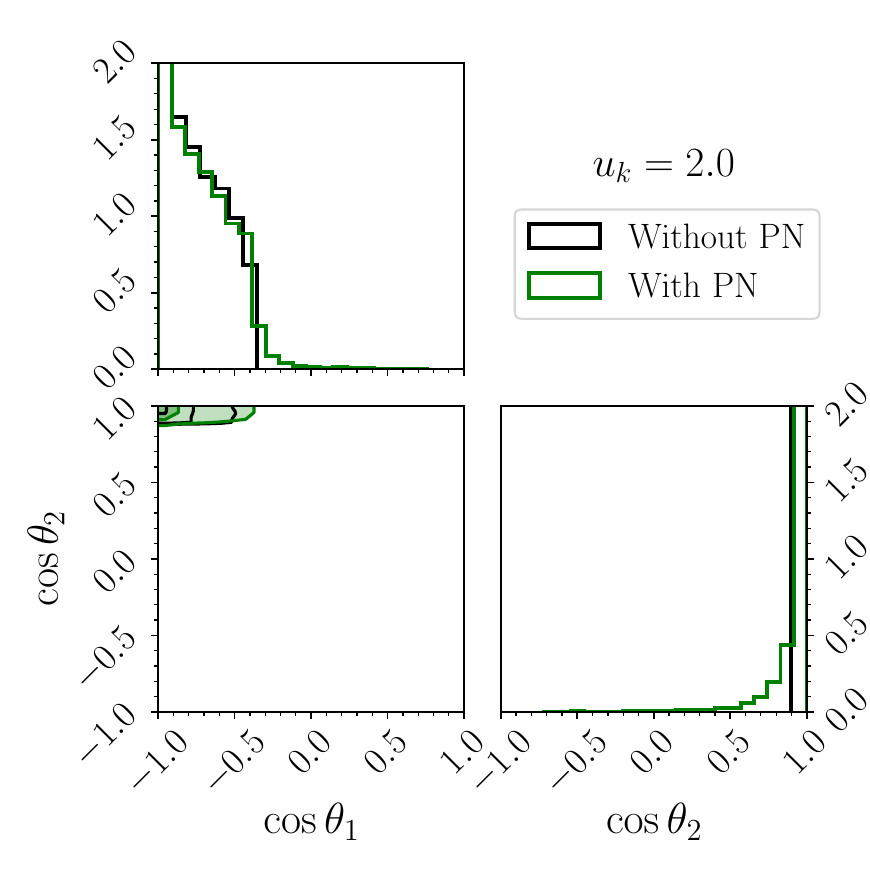}
    \caption{Joint distribution of $\cos\theta_1$ and $\cos\theta_2$ in the field scenario for efficient tides, shown for increasing values of $u_k$. Here we assume that the mass ratio is not reversed during binary evolution, hence the secondary spin $\theta_2$ is close to alignment. Contours refer to the regions containing  $15\%$, $50\%$, and $85\%$ of the probability. Green (black) curves refer to distributions when the PN evolution between BH binary formation and merger is considered (neglected).}
    \label{fig:fieldPN_tides}
\end{figure*}

\begin{figure}
    \includegraphics[width=\columnwidth]{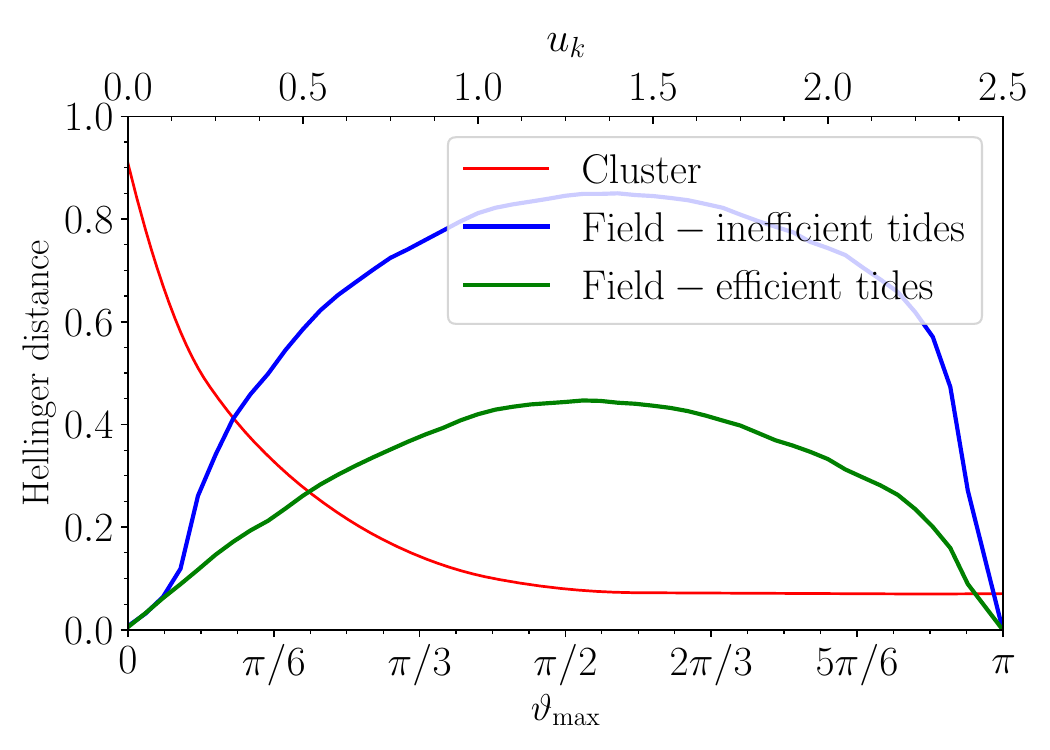}
    \caption{Hellinger distance between the joint distributions of $(\cos\theta_1,\cos\theta_2)$ with and without PN evolution. Red line refers to the cluster 1g+1g binaries while the blue (green) line refers to inefficient (efficient) tides in the field scenario.} 
    \label{fig:hellingerdist1}
\end{figure}

Figure \ref{fig:hellingerdist1} shows the Hellinger distance ---a measure of the distance in probability-distribution space\footnote{The squared Hellinger distance between two distributions $p$ and $q$ is defined as $d^2(p,q) = 1-\int \sqrt{p(x) q(x) } \dd x\in[0,1]$. For instance, the Hellinger distances between identical Gaussian distributions with means separated by $n=1,2,3, 4, 5$ standard deviations are $d\simeq 0.34, 0.63, 0.82, 0.93, 0.98$.
}~\cite{Hellinger+1909+210+271}--- between the joint distributions of $(\cos\theta_1,\cos\theta_2)$ with and without PN evolution, for both formation scenarios. 

In the dynamical case considering the 1g+1g subpopulation, the Hellinger distance decreases as $\vartheta_{\max}$ increases, yielding increasingly similar distributions with and without PN dynamics. This is consistent with earlier results~\cite{2007ApJ...661L.147B,2015PhRvD..92f4016G}, which show that an isotropic distribution of spin directions remains isotropic under PN evolution.

In the isolated case, the distributions are most similar for small and large values of $u_k$, whereas the Hellinger distance is larger for intermediate values. This behavior can be understood in terms of the extent of the initial distribution in the $(\cos\theta_1,\cos\theta_2)$ plane. For intermediate values of $u_k$, the distribution spans a broader region of the parameter space, making PN precession more effective in redistributing the spins over different orientations and therefore leading to a more appreciable modification of the joint distribution. By contrast, for small and large values of $u_k$, the spins are more strongly confined to aligned or anti-aligned configurations, resulting in a smaller effect of PN evolution on the overall distribution. 

When tides are included, the difference between the distributions with and without PN evolution is generally smaller. This is because tidal interactions drive the secondary spin towards alignment with the orbital angular momentum. The reduced extent of the distribution in the $(\cos\theta_1,\cos\theta_2)$ plane limits the redistribution induced by PN spin precession.

\subsection{Impact of the escape speed}\label{finite_esc}

\begin{figure*}
    \centering
    \includegraphics[width=0.33\textwidth]{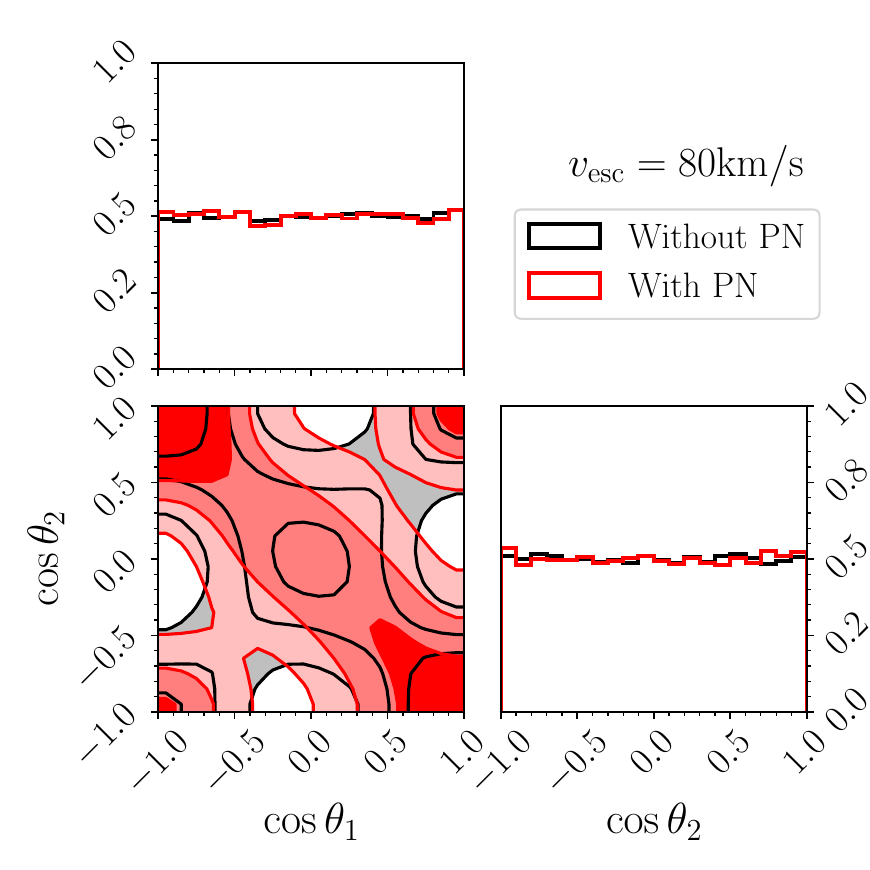}%
    \hfill
    \includegraphics[width=0.33\textwidth]{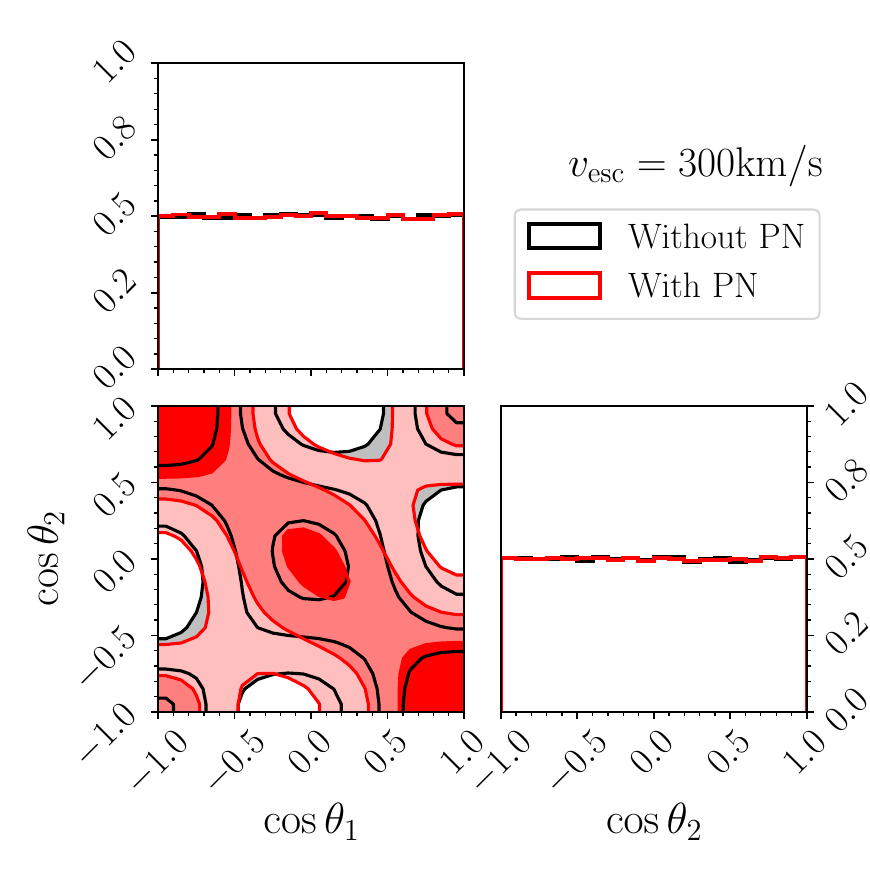}
    \hfill
    \includegraphics[width=0.33\textwidth]{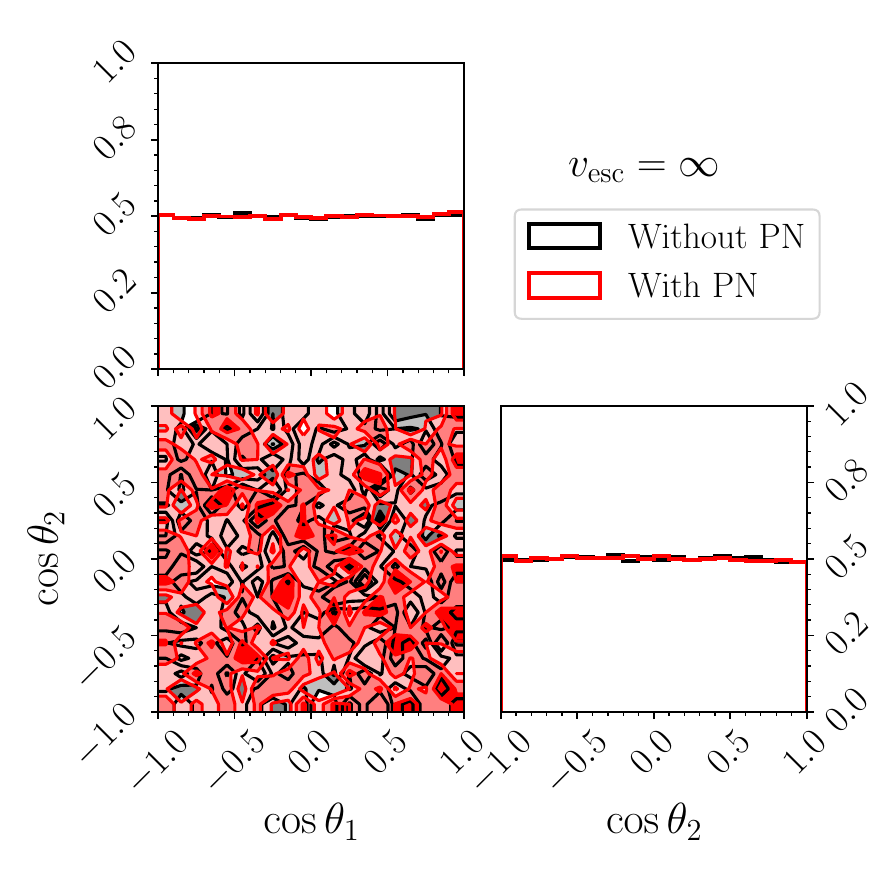}
    \caption{Joint distribution of $\cos\theta_1$ and $\cos\theta_2$ for 1g+2g binaries in the dynamical scenario, shown for increasing values of $v_{\rm esc}$ and assuming $\vartheta_{\rm max}=0$. Contours refer to the regions containing  $15\%$, $50\%$, and $85\%$ of the probability. Red (black) curves refer to distributions when the PN evolution between BH binary formation and merger is considered (neglected). }
    \label{fig:bias12g}
\end{figure*}

In our model for dynamical formation, the spins of BH binaries involving hierarchical mergers are isotropically distributed because they inherit the direction of the angular momentum of the progenitor BH binary, cf. Sec.~\ref{hiermerg}.
Deviations from this behavior are induced by the finite escape speed of the cluster. Figure \ref{fig:bias12g} shows $p(\cos\theta_1,\cos\theta_2)$ for 1g+2g binaries in the extreme case with $\vartheta_{\rm max}=0$, assuming $v_{\rm esc}=80 \rm km/s$, $v_{\rm esc}=300 \rm km/s $ and $v_{\rm esc}=\infty$. For small values of both $\vartheta_{\rm max}$ and $v_{\rm esc}$, an X-shaped structure emerges in the joint distribution. Isotropic distributions are recovered in both limits $v_{\rm esc} \gtrsim 5000$ km/s (this threshold corresponds to the largest BH kicks achievable for quasi-circular BH binaries) and $\vartheta_{\rm max}\lesssim \pi$.

This effect is due to the dependence of BH kicks on the spin orientations. In particular, BH binaries with large spin-orbit misalignments $\theta_{1,2}\sim \pi/2$ and anti-parallel spins are subject to the so-called ``superkick'' effect~\cite{2007PhRvL..98w1102C,2007PhRvL..98w1101G}, ultimately driven by the combined frame dragging of the two Kerr BHs, and receive on average larger kicks than binaries with aligned or anti-aligned spins $\theta_{1,2}\sim 0,\pi$. This implies that BHs that originate from systems whose orbital angular momentum is more closely aligned or anti-aligned with that of the first-generation spins, which in turn correlate with $\vartheta_{\rm max}$, receive smaller kicks and are thus more likely to be retained in the cluster. 

While in principle this effect is also present in the 2g+2g and $>$2g binaries, it is largely washed out in those cases by the numerous random draws involved; in practice, we could not detect it in any of our numerical experiments.

\subsection{Distinguishing field and cluster formation}

Allowing BH spin directions in clusters to exhibit correlations and deviations from isotropy, while allowing spin directions in the field to be misaligned, opens up a discussion on the possibility of distinguishing between the two scenario.

\begin{figure}
    \includegraphics[width=\columnwidth]{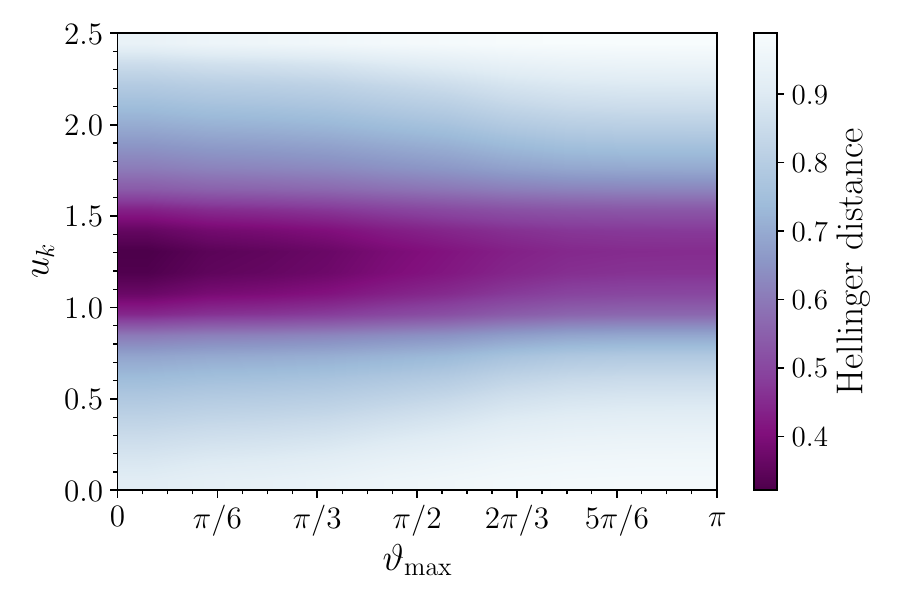}
    \includegraphics[width=\columnwidth]{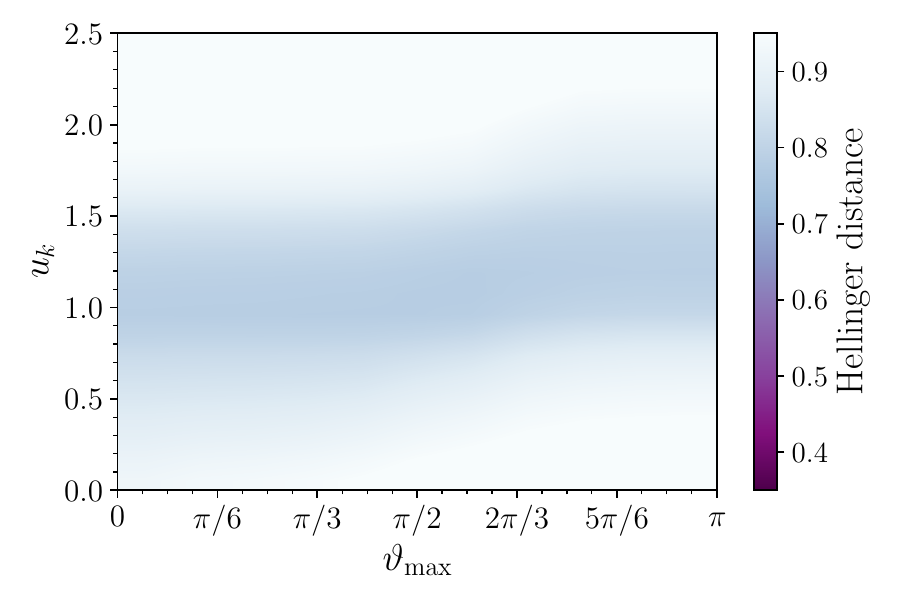}
    \caption{Hellinger distance between the $(\cos\theta_1\cos\theta_2)$ joint distributions for the two formation channels, considering PN evolution and $v_{\rm esc}=80$ km/s. The top (bottom) panel compares dynamical formation against isolated evolution with inefficient (efficient) tides. }
      \label{fig:hellingerdist2d}
\end{figure}

Figure \ref{fig:hellingerdist2d} shows the Hellinger distance between the joint distributions of $(\cos\theta_1, \cos\theta_2)$ for the two formation channels, including PN evolution. For the cluster scenario, we assume a fiducial escape velocity $v_{\rm esc}= 80\rm km/s$, representative of a globular cluster. The resulting fractions of 1g+1g, 1g+2g, 2g+2g and >2g binaries are $\sim 91.9 \%$, $7.5 \%$, $0.5 \%$, and $0.1 \%$, respectively. The top (bottom) panel compares the cluster scenario with the field scenario assuming inefficient (efficient) tides.%

As expected, in the case of inefficient tides, the two distributions are most similar when comparing clusters with small $\vartheta_{\rm max}$ (which maximize deviations from isotropy; cf. Sec.~\ref{clusterresults}) and isolated configurations with intermediate values of $u_k
\gtrsim 1$ (which maximize deviations from alignment; cf. Sec.~\ref{fieldresults}).
The distance between the distributions for dynamical and isolated formation is larger for the case of inefficient tides because the tilt distributions occupy different portions of the available parameter space. 

The cases of fully efficient and fully inefficient tides considered here are idealized and bracket the relevant uncertainties. We argue that the field and cluster distributions remain overall rather different, with $d\gtrsim 0.3$ everywhere in parameter space. While translating this threshold into the number of GW detections required to establish distinguishability requires further work, this is an encouraging finding: even if cluster binaries deviate from isotropy and field binaries deviate from alignment, their spin-orientation distributions remain distinct, corroborating earlier claims~\cite{2016ApJ...832L...2R,2017MNRAS.471.2801S,2017CQGra..34cLT01V,2017PhRvD..96b3012T,2018PhRvD..98h4036G,2023ApJ...946...50B}.

The comparison presented here is necessarily simplified, and several additional aspects should be taken into account. In realistic systems, we expect distributions of both $\vartheta_{\rm max}$ and $u_k$, whereas here we compare idealized scenarios in which those parameters are kept fixed. Moreover, the astrophysical relevance of these parameters may differ, especially in the case of $u_k$. On the one hand, large SN kicks can produce larger spin misalignments; on the other hand, they are also more likely to disrupt the binary. This implies that requiring the system to remain bound preferentially selects systems with more aligned spins, leading to a systematic difference compared to the dynamical scenario. Similarly, in the cluster scenario, hierarchical mergers, whose spins are more isotropically distributed, tend to have more extreme mass ratios in the case of mixed generations and higher masses, potentially beyond GW detectability. Last but not least, let us recall that here we are investigating distinguishability using solely the spin directions, while other observables are routinely used in GW astronomy, notably the mass spectrum, which is better measured than the spin distributions. We plan to investigate correlations between the spin orientations modeled here and other BH parameters in future work.

\section{Conclusions}

In this paper, we have challenged the common expectation that merging BH binaries formed dynamically have spins isotropically distributed, while BH binaries formed in isolation have spins that are aligned with the binary angular momentum and to each other. We have derived analytical distributions for the spin directions in both cases based on simple geometrical considerations, and investigated their robustness with a numerical setup.

Our description of spin directions does not capture the full complexity of the two scenarios considered, including the kinematics and multi-body interactions in dense stellar environments or the detailed evolutionary pathways of binaries formed in isolation. Nevertheless, we argue that this simplified treatment can efficiently capture the key features of the problem while remaining largely independent of the details of the underlying astrophysical models.

The framework presented here is well suited for applications to GW population inference, although some caveats should be kept in mind. In particular, the distributions derived in this work should be weighted by imposing distributions over the parameters $\vartheta_{\rm max}$, $u_k$, $\beta$, $\eta_{\rm hm}$, $\eta_{\rm tides}$, $\eta_{\rm rev}$ that enters our final expressions (\ref{hier_mix}) and (\ref{pfieldall}) for the cluster and field scenario, respectively. In addition, one would need a further parameter $\eta_{\rm ch}$ to model the fraction of binaries in either of the two channels. 
This would results in models of the kind
\begin{align}
&p(\cos\theta_1,\cos\theta_2 | \boldsymbol{\lambda}) \notag\\&\quad= 
p(\vartheta_{\rm max},u_k,\beta, \eta_{\rm hm}, \eta_{\rm tides}, \eta_{\rm rev}, \eta_{\rm ch} | \boldsymbol{\lambda}) 
\notag\\&\quad\times  \big[ \eta_{\rm ch} \, p_{\rm clusters}(\cos\theta_1,\cos\theta_2 | \vartheta_{\rm max}, \eta_{\rm hm})
\notag\\&\quad+(1-\eta_{\rm ch}) \,
p_{\rm field}(\cos\theta_1,\cos\theta_2 | u_k, \beta, \eta_{\rm tides}, \eta_{\rm rev})  \big]\,,
\end{align}
where $\boldsymbol{\lambda}$ refers to a set of hyperparameters to be inferred from GW data.
The functional form of $p(\vartheta_{\rm max},u_k,\beta, \eta_{\rm hm}, \eta_{\rm tides}, \eta_{\rm rev}, \eta_{\rm ch} | \boldsymbol{\lambda}) 
$ can either be chosen flexibly for a more agnostic approach or be directly informed by population-synthesis simulations of the field and cluster scenarios for a more astrophysically targeted analysis.

All predictions presented in this paper refer to the intrinsic properties of BH binaries, which do not take into account GW selection effects. This intrinsic distribution is the typical target of GW population fits~\cite{2026arXiv260527226T,2019MNRAS.486.1086M,2022hgwa.bookE..45V}.
Among our findings, we stress that information about the astrophysical formation pathways of BH binaries may lie in \emph{correlations} between the primary and secondary spin directions, rather than in their marginal distributions. Such correlations are generally not modeled in current GW analyses~\cite{2026arXiv260527226T}, with the recent exception of Ref.~\cite{2026arXiv260910753W}.

More broadly, our results provide a simple analytical framework for connecting binary-formation physics to observable spin-orientation signatures, offering a tractable starting point for testing formation scenarios with current and future GW observations.

\acknowledgements

We thank Giulia Fumagalli, Ilya Mandel, Matthew Mould, Cailin Plunkett, and Alessandro Trani for discussions. %
S.D. is supported by INFN  {\it``More women in Physics''}  scholarship.
S.D., D.G., and T.B. are supported by 
ERC Starting Grant No.~945155--GWmining, 
Cariplo Foundation Grant No.~2021-0555, 
Italian-French University (UIF/UFI) Grant No.~2025-C3-386,
MUR Grant ``Progetto Dipartimenti di Eccellenza 2023-2027'' (BiCoQ),
and the INFN TEONGRAV initiative.
D.G. is supported by MUR Young Researchers Grant No. SOE2024-0000125. %
Computational work was performed 
at CINECA with allocations through INFN and the University of Milano-Bicocca.

\bibliography{notall_notall}

\appendix
\section{Joint spin-spin distribution in the dynamical scenario}
\label{label:jointcluster}

We present the full derivation of Eq.~(\ref{eq:jointcluster}) and discuss some of its properties.  
The quantity of interest, $p(\cos\theta_1, \cos\theta_2)$ is the probability density that two independent directions $\hat{\vec{S}}_1$ and $\hat{\vec{S}}_2$, isotropically distributed in a spherical cap defined by the polar angle range $0 \leq \theta \leq \vartheta_{\mathrm{max}}$, form angles $\theta_1$ and $\theta_2$ with a common direction $\hat{\vec{L}}_{\rm C}$. %

The probability density of having a direction $\hat{\vec S}$ uniformly distributed over a spherical cap of aperture angle $\vartheta_{\mathrm{max}}$ is given by 
\begin{equation}
\label{pScluster}
    p(\hat{\vec S}) =
    \frac{1}{2\pi\left(1-\cos\vartheta_{\mathrm{max}}\right)}
    \,\Theta\!\left(\hat{\vec S}\cdot\hat{\vec L}_{\rm C}-\cos\vartheta_{\mathrm{max}}\right),
\end{equation}
where $\Theta(\cdot)$ denotes the Heaviside step function. %
The probability $p(\hat{\vec S})$ depends only on the scalar product $\hat{\vec S}\cdot\hat{\vec L}_{\rm C}$ and can thus be conveniently expanded using Legendre polynomials:
\begin{equation}
    p(\hat{\vec S})  =\sum_{\ell=0}^{\infty} \frac{2\ell+1}{4\pi}\mu_\ell(\vartheta_{\mathrm{max}})P_\ell(\hat{\vec S}\cdot\hat{\vec L}_{\rm C})\,.
    \label{eq:expansion}
\end{equation}
where the coefficients $\mu_\ell$ are obtained by projecting $p(\hat{\vec S})$ onto the Legendre polynomial basis
\begin{align}
\notag
    \mu_\ell(\vartheta_{\mathrm{max}})
    &=\int p(\hat{\vec S}) P_\ell(\cos\theta')\dd \Omega'\\
    & =
    \frac{1}{1-\cos\vartheta_{\mathrm{max}}}\int_{\cos\vartheta_{\mathrm{max}}}^{1} P_\ell(\cos\theta') \dd\cos\theta'
\end{align}
and $\dd\Omega'=  \sin \theta' \dd\theta' \dd\phi'$ is the solid-angle element. 

The probability density that a direction $\hat{\vec S}$ forms an angle $\theta$ with a given arbitrary direction $\hat{\vec L}_{\rm B}$ is
\begin{equation}
    p(\cos\theta|\hat{\vec L}_{\rm B})=\int p(\hat{\vec S})\delta(\cos\theta'-\cos\theta)\dd \Omega'\,.
    \label{eq:condition}
\end{equation}
Plugging Eq.~(\ref{eq:expansion}) into Eq.~(\ref{eq:condition}) and using the properties of Dirac's delta yields 
\begin{align}
    p(\cos\theta|\hat{\vec L}_{\rm B})=
    &\sum_{\ell=0}^{\infty} \frac{2\ell+1}{4\pi}\mu_\ell(\vartheta_{\textrm{max}}) 
   \notag \\
   & \times
    \int_0^{2\pi} P_\ell(\hat{\vec S}\cdot\hat{\vec L}_{\rm C})\big|_{\cos\theta=\cos\theta'}\dd \phi'\,.
\end{align}
The integral in the equation above can be computed by applying properties of Legendre polynomials
\begin{align}
\notag    &\int_0^{2\pi} P_\ell(\cos\theta\cos\alpha+\sin\theta\sin\alpha\cos\phi')\dd \phi'\\
    &
    \quad= 2\pi P_\ell(\cos\theta)P_\ell(\cos\alpha)\,, 
\end{align}
where $\hat{\vec S}\cdot\hat{\vec L}_{\rm C}$ has been expressed in terms of coordinates around $\hat{\vec L}_{\rm B}$ such that $\cos\alpha = \hat{\vec L}_{\rm B}\cdot\hat{\vec L}_{\rm C}$. %
The result is
\begin{align}
    p(\cos\theta|\hat{\vec L}_{\rm B})
    &=\sum_{\ell=0}^{\infty}\frac{2\ell+1}{2}\mu_\ell(\vartheta_{\textrm{max}})
    P_\ell(\cos\theta)P_\ell(\cos\alpha)\,. 
    \label{eq:condfinal}
\end{align}

For a fixed direction $\hat{\vec L}_{\rm B}$, the joint probability density that the directions $\hat{\vec S}_1$ and $\hat{\vec S}_2$ form angles $\theta_1$ and $\theta_2$ with $\hat{\vec L}_{\rm B}$ factorizes into the product of the individual probability densities [Eq.~(\ref{eq:condfinal})], since $\hat{\vec S}_1$ and $\hat{\vec S}_2$ are statistically independent:
\begin{equation}
    p(\cos\theta_1, \cos\theta_2|\hat{\vec L}_{\rm B})=p(\cos\theta_1|\hat{\vec L}_{\rm B})\,p(\cos\theta_2|\hat{\vec L}_{\rm B})\,.
\end{equation}
Finally, $p(\cos\theta_1,\cos\theta_2)$ is obtained by averaging over all possible directions $\hat{\vec L}_{\rm B}$ with $p(\cos\alpha)=1/2$, resulting in
Eq.~(\ref{eq:jointcluster}). 

The marginal distributions $p(\cos\theta_1)$ and $p(\cos\theta_2)$ are both uniform, as they describe the angle between a direction confined to a spherical cap and an isotropically distributed direction. This result is straightforward to verify analytically:
\begin{align}
    \notag
    p(\cos\theta_1)
    &=\int_{-1}^{1}p(\cos\theta_1,\cos\theta_2)\dd (\cos\theta_2)=\\
    \notag
    &=\sum_{\ell=0}^{\infty}\frac{2\ell+1}{4}\mu_\ell^2(\vartheta_{\text{max}})P_\ell(\cos\theta_1) \times 2\delta_{l0}\\
    \notag
    &=\frac{1}{2}\mu_0^2(\vartheta_{\text{max}})P_0(\cos\theta_1)=\frac{1}{2}\,. 
\end{align}
The proof is identical for $p(\cos\theta_2)$.

\section{Supernova spin tossing}  \label{tossapp}

In this section, we generalize our model for spin directions in cluster to the case where the SN explosion impart another tilt to the spin of the newly formed BH, elsewhere referred to spin tossing. 
Let us take a spin direction $\hat{\vec S}$ and impart a tilt with largest amplitude $\vartheta_{\rm toss}$, uniformly distributed in cosine. In full analogy with the calculation presented in Appendix~\ref{label:jointcluster}, the probability of the post-toss spin direction ${\hat{\vec S}}'$ is given by
\begin{align}
    p({\hat{\vec S}}'  | \hat{\vec S}) &=
    \frac{1}{2\pi\left(1-\cos\vartheta_{\mathrm{toss}}\right)}
    \,\Theta\!\left({\hat{\vec S}}'\cdot\hat{\vec S}-\cos\vartheta_{\mathrm{toss}}\right)
  \notag \\&=\sum_{\ell=0}^{\infty} \frac{2\ell+1}{4\pi}\mu_\ell(\vartheta_{\mathrm{toss}})P_\ell({\hat{\vec S}}'\cdot\hat{\vec S})\,.
\end{align}
The total probability of reaching direction $\hat{\vec S}'$ is obtained marginalizing over $\hat{\vec S}$
\begin{equation}
    p(\hat{\vec S}') = \int p(\hat{\vec S}'|\hat{\vec S})p(\hat{\vec S})d\hat{\vec S}\,,
\end{equation}
where $p(\hat{\vec S})$ is given by Eq.~(\ref{eq:expansion}). Some straightforward algebra leads to
\begin{equation}
    p(\hat{\vec S}') = \sum_{\ell=0}^{\infty} \frac{2\ell+1}{4\pi}\mu_\ell(\vartheta_{\mathrm{max}})\mu_\ell(\vartheta_{\mathrm{toss}})P_\ell(\hat{\vec S}'\cdot\hat{\vec L}_{\rm C})\,,
\end{equation}
and
\begin{equation}
    \begin{split}
p(\cos\theta_1,\cos\theta_2)
&=\sum_{\ell=0}^{\infty}
\frac{2\ell\!+\!1}{4}\,
[\mu_\ell(\vartheta_{\text{max}})\mu_\ell(\vartheta_{\text{toss}})]^2\\
&\times
P_\ell(\cos\theta_1)
P_\ell(\cos\theta_2)\,,
    \end{split}
\end{equation}
Compared to the model presented in the main body of the paper, including an additional tilt due to SN spin tossing amounts to introducing a second $\mu_l$ coefficient in Eq.~(\ref{eq:jointcluster}). The resulting model is therefore symmetric in the $(\vartheta_{\rm max},\vartheta_{\rm toss})$ parameter space under the exchange $\vartheta_{\rm max}\leftrightarrow\vartheta_{\rm toss}$.

\begin{figure*}
    \centering
    \includegraphics[width=0.31\textwidth]{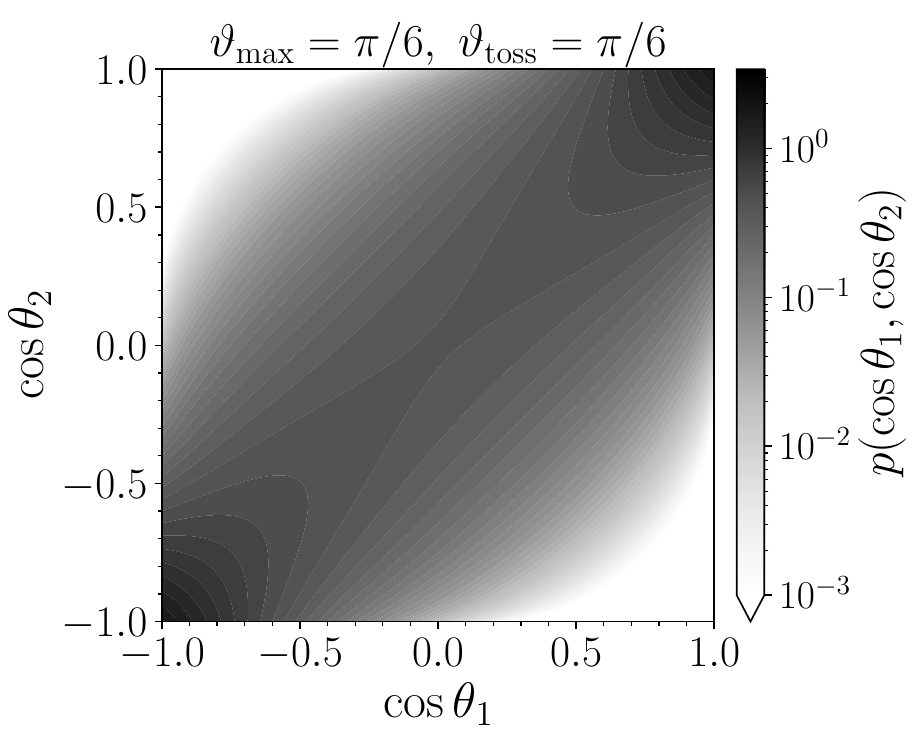}
    \hfill
    \includegraphics[width=0.31\textwidth]{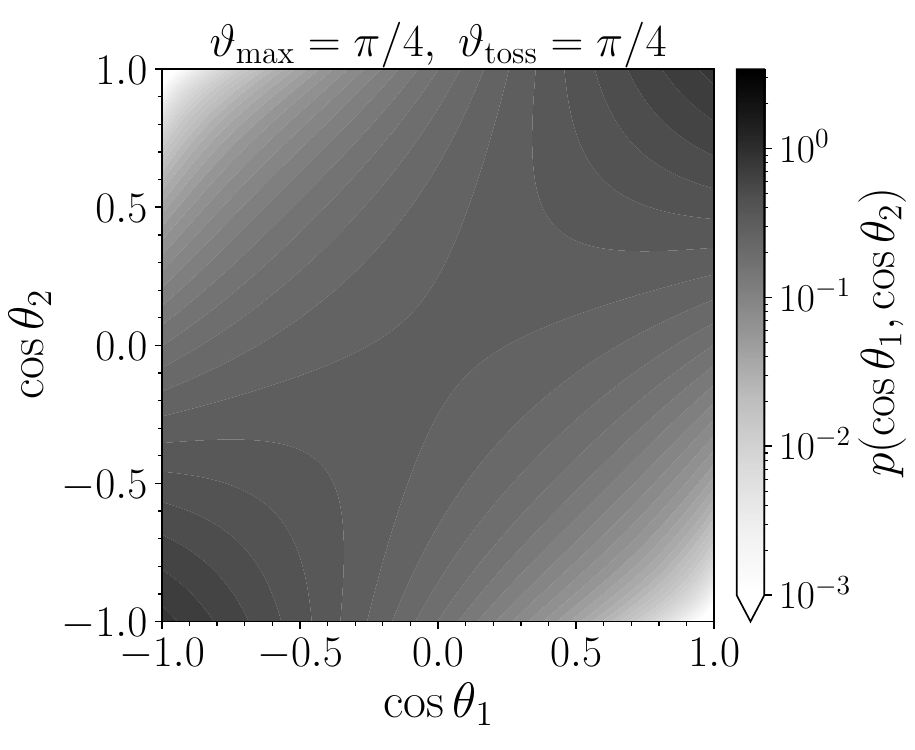}
    \hfill
    \includegraphics[width=0.31\textwidth]{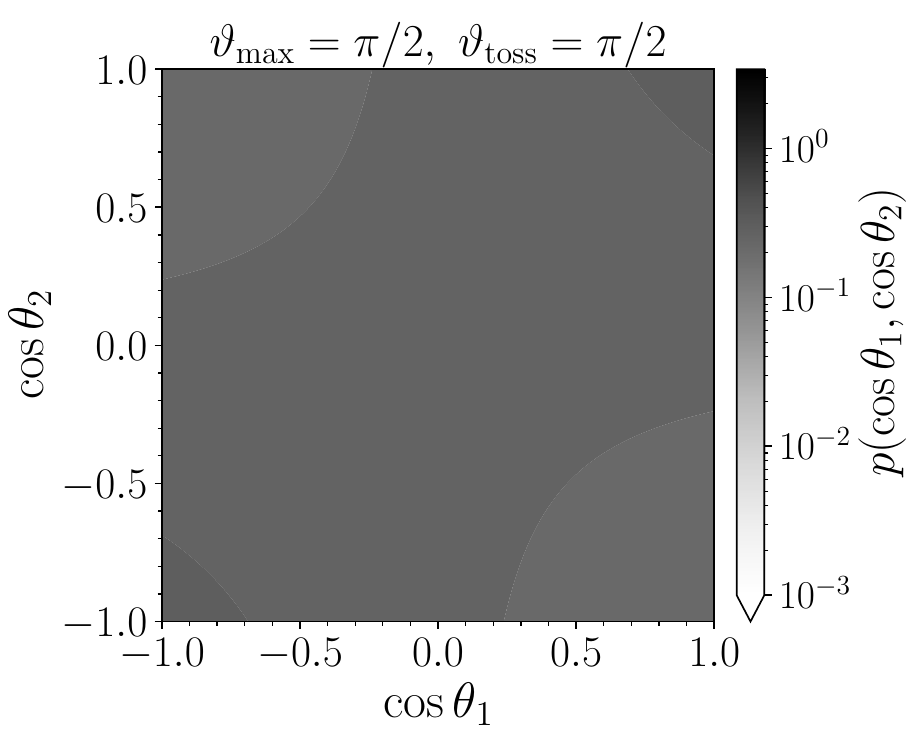}
    \caption{Joint distribution $p(\cos\theta_1,\cos\theta_2)$ of the two spin tilt angles for first-generation binary BHs in clusters, allowing for SN spin tossing. In each panel, we show possible combinations of the angles $\vartheta_{\rm max}$ and $\vartheta_{\rm toss}$.}
    \label{fig:toss}
\end{figure*}

Figure \ref{fig:toss} shows the joint distribution $p(\cos\theta_1,\cos\theta_2)$ for a few  different combinations of $\vartheta_{\rm max}$ and $\vartheta_{\rm toss}$. 

\section{Distribution of the orbital plane tilt in isolated binaries}
\label{label:anglefield}

In this section, we provide the full derivation of Eqs.~(\ref{eq:fieldpdf})-(\ref{intervalrho}). Our starting point is the expression for the orbital-plane tilt in Eq.~(\ref{eq:cosg}), together with the bound-system condition given by Eq.~(\ref{eq:survivecond}).

The kick direction is given by $\hat{\rm \vec{v}}_k =(v_{k,x},v_{k,y}, v_{k,z})=(\sin\theta_k\cos\phi_k,\sin\theta_k\sin\phi_k,\cos\theta_k)$, where we define a Cartesian triad with the $z$-axis along the orbital velocity of the exploding star, the $x$-axis aligned with the pre-SN orbital angular momentum, and the $y$-axis along the instantaneous binary separation vector. %
Assuming that the pre-SN orbit is circular, that kicks are isotropic and conditioning on binaries that remain bound, one has $p(\phi_k)=\mathcal{U}[0,2\pi]$ and $p(\cos\theta_k)=\mathcal{U}[-1,\cos\theta_{k,\text{max}}]$.

The probability density of the direction $\hat{\vec{v}}_{\rm k}$ is then:
\begin{equation}
    p(\rm \hat{\vec{v}}_k)=\frac{1}{2\pi(\cos\theta_{k,\text{max}}+1)}\,. 
    \label{fdist}
\end{equation}

Defining $\omega_1=1+u_k \cos\theta_k$ and $\omega_2=u_k \sin\theta_k \cos\phi_k$, Eq.~(\ref{eq:cosg}) can be written as
\begin{equation}
    \cos\gamma=\frac{\omega_1}{\sqrt{\omega_1^2+\omega_2^2}}\,.
\end{equation}
This implies that finding the distribution of $\gamma$ is equivalent to determining the polar-angle distribution of the vector $\boldsymbol{\omega}=(\omega_1,\omega_2)$. Since $(\omega_1-1)^2+\omega_2^2\leq u_k^2$, the tip of $\boldsymbol{\omega}$ must lie within a disk of radius $u_k$ centered at $(1,0)$. However, the allowed region is not the full disk because $-1\leq \cos\theta_k\leq \cos\theta_{k,\text{max}}$, which implies $1-u_k\leq \omega_1\leq 1+u_k\cos\theta_{k,\text{max}}$. The length of the allowed interval in $\omega_1$ is therefore $u_k(\cos\theta_{k,\text{max}}+1)$.

Although $\rm \hat{\vec{v}}_k$ is uniformly distributed on the sphere, the resulting distribution of $\boldsymbol{\omega}$ is not uniform on its plane. 
To derive its distribution, one must first project $\hat{\rm\vec{v}}_k$ onto the spherical surface spanned by $(v_{k,x},v_{k,z})$ and then perform a change of variables to $(\omega_1,\omega_2)$.
This leads to
\begin{align}
    \notag
    p(v_{k,x},v_{k,z})
    &=\!\frac{1}{\pi(\cos\theta_{k,\text{max}}\!+\!1)}\\
    &
    \times\frac{1}{\sqrt{1\!-\!\sin^2\!\theta_k \cos^2\!\phi_k\!-\! \cos^2\!\theta_k} 
    }
\end{align}
and 
\begin{align}
    \notag
    p(\omega_1,\omega_2)
    &=\frac{1}{\pi(\cos\theta_{k,\text{max}}+1)u_k}\\
    &
    \times \frac{1}{\sqrt{u_k^2-(\omega_1-1)^2-\omega_2^2}}, 
\end{align}
with the constraints: 
\begin{enumerate}
    \item $(\omega_1-1)^2+\omega_2^2\leq u_k^2$.
    \item $1-u_k\leq \omega_1\leq 1+u_k\cos\theta_{k,\text{max}}$.
\end{enumerate}   

We now write $\boldsymbol\omega$ using polar coordinates
\begin{equation}
    \omega_1=\rho\cos\gamma\,, \qquad \omega_2=\rho\sin\gamma\,,
\end{equation} 
 where $\rho\geq 0$ and $\gamma$ is the variable of interest. Integrating over $\rho$ yields
\begin{equation}
    p(\gamma)=\int_{\mathcal{I}}p(\rho\cos\gamma,\rho\sin\gamma)\rho\dd \rho\,,
    \label{interval}
\end{equation}
where $\mathcal{I}$ is the interval of the integration allowed by the constraints above. In particular:
\begin{enumerate}
    \item The first constraint is $\rho^2-2\rho\cos\gamma+1\leq u_k^2$. The discriminant of the quadratic inequality is $h(\cos\gamma)=\sqrt{u_k^2+\cos^2\gamma-1}$. If $u_k^2+\cos^2\gamma-1\leq 0$, there are no valid solutions. Otherwise, one has  $|  \rho - \cos\gamma | \leq h(\cos\gamma) $
    \item The second constraint is $1-u_k\leq \rho \cos\gamma \leq 1+u_k\cos\theta_{k,\text{max}}$.
\end{enumerate}  
Putting all the constraints together, the interval $\mathcal{I}$ of permitted values of $\rho$ is that given by Eq.~(\ref{intervalrho}).
From Eq.~(\ref{interval}) one gets
\begin{align}
    \notag
    p(\gamma)
    &=\frac{1}{\pi (\cos\theta_{k,\text{max}}+1)u_k}\\
    &
    \times\int_{\mathcal{I}}\frac{\rho\dd \rho}{\sqrt{h^2(\cos\gamma)-(\rho-\cos\gamma)^2}}\,,
\end{align}
and thus
\begin{align}
   \notag
    p(\cos\gamma)
    &=
    \frac{2}{\pi (\cos\theta_{k,\text{max}}+1)u_k\sqrt{1-\cos^2\gamma}}\\
    &
    \times\int_{\mathcal{I}}\frac{\rho\dd \rho}{\sqrt{h^2(\cos\gamma)-(\rho-\cos\gamma)^2}}\,. 
\end{align}
The integral above can be carried out analytically, leading to the antiderivative reported in Eq.~(\ref{antiderivative}), and thus
 to the final expression of  Eq.~(\ref{eq:fieldpdf}). %

\end{document}